\documentclass[fleqn,usenatbib]{mnras}

\usepackage{newtxtext,newtxmath}

\usepackage[T1]{fontenc}

\DeclareRobustCommand{\VAN}[3]{#2}
\let\VANthebibliography\thebibliography
\def\thebibliography{\DeclareRobustCommand{\VAN}[3]{##3}\VANthebibliography}

\usepackage{graphicx}	
\usepackage{amsmath}	
\DeclareUnicodeCharacter{2212}{-}
\usepackage{comment}
\usepackage{xcolor}

\title[Wavelet analysis of AGN and QPEs]{Applying wavelet analysis to the X-ray light curves of active galactic nuclei and quasi-periodic eruptions}

\author[A. Ghosh et al.]{
Akshay Ghosh\thanks{E-mail: akshay.ghosh@smu.ca},
L. C. Gallo,
A. G. Gonzalez
\\
Department of Astronomy and Physics, Saint Mary’s University, 923 Robie Street, Halifax, NS, B3H 3C3, Canada
}

\date{Accepted XXX. Received YYY; in original form ZZZ}

\pubyear{2023}

\begin{document}
\label{firstpage}
\pagerange{\pageref{firstpage}--\pageref{lastpage}}
\maketitle

\begin{abstract}
In this work, we examine the application of the wavelet transform to the X-ray timing analyses of active galactic nuclei (AGN) and quasi-periodic eruption sources (QPEs).   Several scenarios are simulated to test the effectiveness of the wavelet analysis to stationary and non-stationary data.   We find that the power spectral density (PSD) slope and the nature of the periodic signal can influence the ability to identify important features in the wavelet power spectrum.   In general, weak and transient features can be discerned, which make the wavelet spectrum an important tool in examining AGN light curves.  We carried out a wavelet analysis to four unique objects: Ark 120, IRAS~13224-3809, RE J1034+396, and the QPE GSN 069.  The well-known quasi-periodic oscillation (QPO) in RE J1034+396 is significantly detected in the wavelet power spectrum.  In IRAS~13224-3809, significant transient features appear during a flare at frequencies coincident with previously detected reverberation signals.  Finally, the wavelet power spectrum of the QPE GSN 069 significantly reveals four persistent signals that exhibit a 3:2 ratio in oscillation frequencies, consistent with high-frequency QPOs in stellar mass X-ray binaries, but we cannot rule out the possibility this is an artefact of the calculation.
\end{abstract}




\section{INTRODUCTION}

Active galactic nuclei (AGN) consist of an accretion disk surrounding a host supermassive black hole (SMBH).  The X-rays originate from a compact corona of hot electrons located close to the black hole.  X-ray radiation from AGN can be extremely variable on all observable timescales, from hours (e.g. \citealt{vaughan2003characterizing,paolillo2004_shorttermvar,gallo2018_shorttermvar}) to years (e.g. \citealt{markowitz2001rxte_longtermvar,gonzalezmartin2012}). This variability provides insight to the physical processes that power these sources (e.g. \citealt{vaughan2003characterizing, uttley2014}).  Current analyses of this variability are usually based on Fourier techniques like the power spectral density (PSD) and lag-frequency analysis. Fourier techniques operate in the frequency domain and provide the average frequency distribution throughout the entire time series, but there is no time localization. 

The Fourier transform assumes a stationary time series.  For a stationary X-ray light curve, the count rate will be log-normally distributed and exhibit a linear rms-flux relation (\citealt{uttley2005nonlinear,alston2019non_stationary}).  However, there is evidence to suggest that AGN emission is actually non-stationary (e.g. \citealt{gliozzi2004nonstationary, alston2019non_stationary}). The defining characteristic of a stationary time series is that all of its statistical moments are constant (strong stationarity). A weakly stationary time series implies only the mean and variance remain constant (the first and second statistical moments). However, because of the scatter in the statistical moments that are intrinsic to red noise stochastic processes, different realizations of the same process can look very different from one observation to another, even with identical underlying properties (e.g. the PSD and probability distribution function [PDF]). In a time series analysis, we would like to infer the underlying physical process and not a single instance of that process which is affected by its stochastic nature (whether stationary or not), therefore it is useful to examine the time dependence of the expectation values of the statistical moments \citep{vaughan2003characterizing}. 

Physical changes to the AGN such as changes to the accretion disk or corona geometry could lead to non-stationarity in the flux variability (e.g. \citealt{alston2019remarkable}, \citealt{panagiotou2022_uv_xray_correlation}). In general, a non-stationary process could be attributed to the properties generating that process being dynamic.\

\citet{panagiotou2022_uv_xray_correlation} set out to explain the recent trend of lower-than-expected correlations between UV and X-ray emissions from the same object, where a correlation is expected when UV reprocessed emission from the accretion disk is driven by the X-ray illumination of the disk from the Comptonized photons of the corona. They discuss how the cross-correlation function assumes that the input time series are stationary, as this is what causes a constant lag between the time series. In their simulations, the reprocessed UV emission depends on the X-ray emission and a response function which encodes the processes taking place in the accretion disk. The input parameters for the simulation correspond to the geometry and physical state of the AGN. More specifically, the height of the corona above the plane of the accretion disk (geometry), and the power law for the X-ray energy spectrum for the corona (physical state). For a static system, both of these parameters are held constant, and from this a stationary process would arise as there would be a single response function governing the reprocessing. However, when the system is dynamic, i.e. either the geometry or physical state have some time dependence, there must be a set of response functions, each corresponding to a point in the parameter space. Not only would this cause a lower-than-expected cross-correlation, it would also cause the AGN emission to be non-stationary.

Wavelet analysis, and more specifically the wavelet transform, can be thought of as an extension to the Fourier transform where the frequency decomposition can be carried out as a function of time.  Unlike in the Fourier transform where the time information is lost, the wavelet analysis combines the time and frequency domains. In the most general sense, there are two overall types of wavelet transforms: the discrete wavelet transform (DWT), and continuous wavelet transform (CWT). The DWT is typically used for ``practical'' applications such as image processing (e.g. \citealt{broughton1998wavelet_dwt_imageprocessing}, \citealt{chang2007direction_dwt_imageprocessing}, \citealt{chervyakov_dwt_imageprocessing}); communications (e.g. \citealt{akansu1998orthogonal_dwt_comms}, \citealt{saad2010_dwt_comms}, \citealt{baig2018_dwt_comms}); and data compression, especially in the medical field (e.g. \citealt{badawy2002mri_dwt_datacomp}, \citealt{qu2003_dwt_datacomp}, \citealt{jha2021empirical_dwt_datacomp}). The CWT is typically used for data analysis and scientific research such as seismology (e.g. \citealt{li2009evaluation_cwt_seis}, \citealt{karamzadeh2012_cwt_seis}, \citealt{balafas2015_cwt_seis}); medicine (e.g. \citealt{bostanov2004bci_cwt_med}, \citealt{cheng2010_cwt_med}, \citealt{komorowski2016_cwt_med}); finance (e.g. \citealt{kristoufek2013_cwt_finance}, \citealt{olayeni2016_cwt_finance}, \citealt{tiwari2016_cwt_finance}); and understanding the impact of COVID-19 on the stock market (e.g. \citealt{caferra2021_cwt_stocks}, \citealt{goodell2021covid_btc_cwt_stocks}, \citealt{umar2022covid_cwt_stocks}).

In general astronomy, wavelet analysis has been used to study chromospheric variations in main sequence stars \citep{frick1997wavelet}; connecting different solar periodicities to a common physical mechanism \citep{krivova2002_solar_cwt}; deviations in the vertical structure of the outer Galactic HI disk \citep{levine2006_vertical_galactic_HI_cwt}; quasi-periodic pulsations of solar flares \citep{dominique2018detection_solar_flares_cwt}; short-lived, narrow-banded solar emission peaks \citep{suresh2017wavelet}; and searching for periodicities in the optical light curves of the blazar S5 0716+714 \citep{gupta2008periodic_optical}.

In the study of AGN and compact objects, wavelet analysis has been used to study the characteristic timescales (radio frequencies) of a large sample of AGN \citep{hovatta2008wavelet}; the X-ray variability of ultraluminous X-ray sources over large timescales \citep{lin2015longterm_ulxs}; constraining the temperature of the intergalactic medium using high-redshift quasars \citep{wolfson2021_igm_temp_cwt}; transient quasi-periodic oscillations (QPOs) in low-mass X-ray binaries \citep{chen2022wavelet}; the search for QPOs in the $\gamma$-ray light curves of blazars \citep{ren2022quasi}; characterizing the X-ray flickering of cataclysmic variables \citep{anzolin2010wavelet}; and the possible detection of a QPO in the narrow-line Seyfert 1 galaxy MCG–06–30–15 \citep{gupta2018_qpo}.

Wavelet analysis is relatively underused in the field of AGN X-ray astronomy. In this paper we will examine the application of the wavelet transform power spectrum on the X-ray light curves of AGN and quasi-periodic eruption (QPE) sources.  In Section 2, we discuss the specifics of wavelet transforms. In Section 3, we discuss the simulations used for testing and in Section 4 we apply the wavelet analysis to four AGN systems.  We discuss and summarise our results in Section 5.

\section{WAVELET TRANSFORMS}

The wavelet transform is similar in essence to the Fourier transform as it calculates the frequencies present in a time series. However, there is a key difference. The Fourier transform is averaged over the entire series and all time domain information is lost.  On the other hand, the wavelet transform preserves the timing information and therefore can be used to identify when specific frequencies appear in the time series. For a comprehensive review of wavelet transforms, see \cite{torrence1998} and the textbook by \cite{Addison}.

Similar to how the Fourier transform expresses the time series as a  sum of sines and cosines, the wavelet transform uses basis vectors known as wavelets or the mother wavelet, denoted by $\psi(t)$. The wavelet undergoes a sequence of two different transformations, one in translation and the other in dilation. The combination identifies the different frequencies that are prominent at each point in time.

The wavelet transform does not assume a stationary time series (for a review of stationarity, see \citealt{alston2019remarkable}) and, in principle, can be used on data with uneven sampling (e.g. \citealt{foster1996wavelets}, \citealt{daubechies1999wavelets}, \citealt{bravo2014wavelets}).  Consequently, wavelet analysis should be ideal to study AGN X-ray light curves.  
For this work, computations of the continuous wavelet transform are carried out using the {\sc matlab} application called {\sc wavelet toolbox}.


The wavelet transform $T(a,b)$ of a time series $x(t)$ is defined as the integral over all times of the series multiplied by the complex conjugate of the wavelet basis vector \citep{Addison}:
\begin{equation}
    T(a,b) = w(a)\int_{-\infty}^{+\infty} x(t)\psi^{*}\left( \frac{t - b}{a} \right) dt, \label{wavelet_eqn}
\end{equation}
where $w(a)$ is the weighing function. In the {\sc wavelet toolbox}, $w(a)$ is chosen such that it uses L1 normalization, which preserves wavelet power spectrum amplitudes at different frequencies \citep{misiti1996wavelet}. In other words, with the L1 normalization scaling, the time series $y(t) = A_i \sin (f_i t) + A_j \sin(f_j t)$, will have the wavelet power spectrum amplitudes proportional as $A_i/A_j$ for any $f_i$ and $f_j$.

The parameters $a$ and $b$ represent the scaling and shifting of the wavelet basis vector $\psi$, respectively. The scaling parameter is similar to the scaling in the Fourier transform, which represents each frequency component. Therefore $a$ can be considered equivalent to the Fourier frequency. The shifting parameter is not present within the Fourier transform and represents the time information that is lost when taking a Fourier transform.  Therefore, the shift parameter $b$ can be considered as the time. In practice, the wavelet transform takes on the form $T(t,f)$, that is, a function of time and frequency. For $\psi = e^{i2\pi t/a}$, Eq.~\ref{wavelet_eqn} reduces to the Fourier transform.

Typically, the wavelet transform is calculated using a convolution between $x$ and the Fourier transform of $\psi$, denoted as $\Psi$. The convolution theorem states that the Fourier transform of the convolution of two vectors is equal to the products of their respective Fourier transforms. This is useful, because convolution takes care of the shifting aspect of the wavelet transform.

Like the Fourier transform, the wavelet transform is not restricted to the real numbers, and can exist over the complex numbers. Similar to how the PSD of a time series is defined by the modulus square of the Fourier transform  \citep{uttley2014},  the wavelet power spectrum, $W(t,f)$, is defined as the modulus square of the wavelet transform \citep{torrence1998}:
\begin{equation}
W(t,f) = |T(t,f)|^2.
\end{equation}

When calculating the wavelet power spectrum of a time series, a wavelet basis vector (mother wavelet) must be defined. While this step may appear arbitrary and could arguably lead to differing results, in essence it is no different with the Fourier transform because the choice of the basis functions are the sine and cosine functions (i.e. complex exponential).  The wavelet basis functions are typically orthogonal and complex. 

A commonly used wavelet basis is the Morlet wavelet, which is a complex exponential multiplied by a Gaussian window. One useful aspect of the Morlet wavelet is its symmetry with the Fourier transform, which is a complex exponential spanning all $t$.
The Morlet wavelet at a central frequency of $f_0$ is defined by \citet{torrence1998} as:
\begin{equation}
    \psi(t,f) = \pi^{-1/4} e^{i2\pi f_0 t} e^{-t^2/2}.
\end{equation}

Another commonly used wavelet basis vector is the generalized Morse wavelet \citep{lilly2008_morse_wavelets}, which is defined in the time domain as:

\begin{align}
    \psi_{P,\gamma} (t) = \frac{1}{2\pi} \int_{0}^{\infty} a_{P,\gamma} f^{\frac{P^2}{\gamma}} e^{-f\gamma} e^{i f t} df.
\end{align}

The generalized Morse wavelet is parameterized by the time-bandwidth ($P$) and symmetry ($\gamma$). The parameter $P$ controls the time/frequency variance. A low value results in a wavelet that is more localized in time and spread out in frequency, while a high value of $P$ results in a wavelet that is more localized in frequency and spread out in time. In the {\sc wavelet toolbox}, this parameter can exist in the range $P \in [\gamma,120]$. 

The $\gamma$ parameter dictates the symmetry of the wavelet about the central frequency $f_c = (P/\gamma)^{2/\gamma}$. The wavelet is most symmetric when $\gamma = 3$ \citep{lilly2008_morse_wavelets}. For this work, $\gamma$ is held constant at 3, and $P = 60$, which is the {\sc matlab} default value, was used for each wavelet power spectra unless specified otherwise. The parameter $a_{P,\gamma}$ is a normalization constant.


The translation and dilation of the wavelet basis vector is ultimately what identifies the prominent frequencies at points in time. The wavelet power spectrum can essentially be thought of as the modulus square of the inner product between the time series and the wavelet basis vector at each time--frequency pairing, up to some resolution.

\section{SIMULATIONS}

To test the effectiveness of wavelet analysis for AGN X-ray light curves, several investigations are carried out on simulated light curves.  Before presenting specific simulations, we describe how the light curves are generated and how the wavelet transform can be examined.

\subsection{Method for simulating AGN X-ray light curves} \label{lc_sim_section} 

The simulations are based on the method of \cite{emmanoulopoulos2013generating}, which uses an iterative method to build off the classic light curve simulations of  \cite{timmer1995generating}, which generates Gaussian distributed light curves for a given PSD. \cite{emmanoulopoulos2013generating} uses phase and amplitude adjustments to modify the \citet{timmer1995generating} light curves, such that their flux can be described by some specified distribution. This better replicates AGN X-ray light curves, which typically have fluxes that are log-normally distributed \citep{uttley2005nonlinear}.

The three input parameters in the \citet{emmanoulopoulos2013generating} method are the light curve flux distribution, PSD, and number of iterations. For the flux distribution, if a simulated light curve was meant to replicate an observation, the distribution of that observation was adopted; otherwise, a theoretical log-normal distribution was used. For the PSD, a simple power law functional form of $P\left(f\right) \propto f^{-\beta}$, where $\beta$ is the power law slope, was used, with the option to add in one or multiple Lorentzian profiles to replicate QPOs. For each simulation, 80 iterations were used as the process typically converges after $\sim 50$ iterations.

Due to the counting nature of AGN X-ray emission, these light curves are a Poisson process and therefore contain a Poisson noise component. 
The Poisson noise component is added to the simulated light curve $x_{sim}$ by resampling each point at times $t_i$ from a Poisson distribution based on the bin size $\Delta t$ \citep{emmanoulopoulos2013generating}:

\begin{align}
    x(t_i) = \frac{Pois(\mu = x_{sim}(t_i)\Delta t)}{\Delta t},
\end{align}
where $\mu$ is the mean of the Poisson distribution being sampled. All of the observed and simulated light curves in this work have a bin size of 100 s.

\begin{figure*}
    \centering
    \includegraphics[width=\columnwidth]{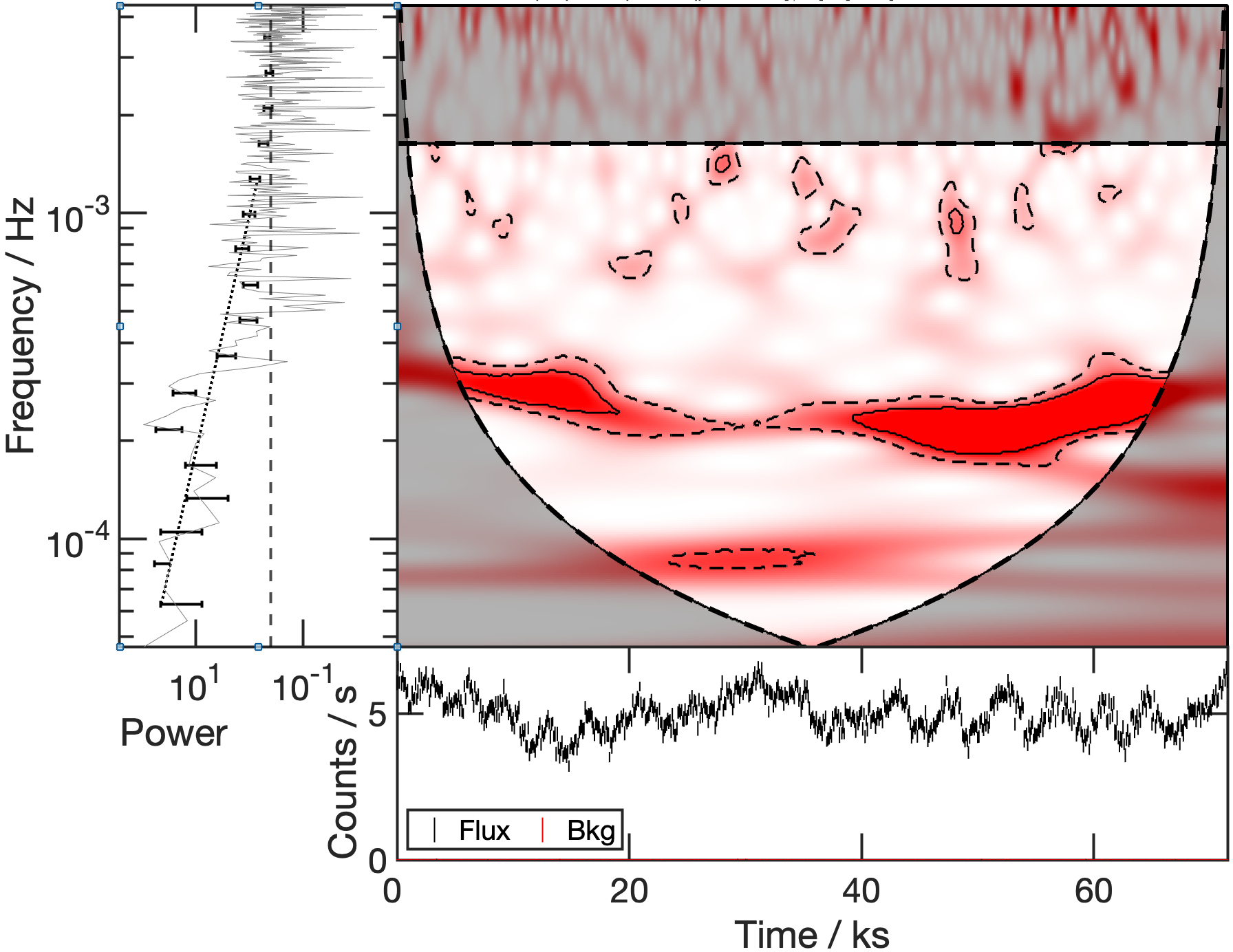}
    \includegraphics[width=\columnwidth]{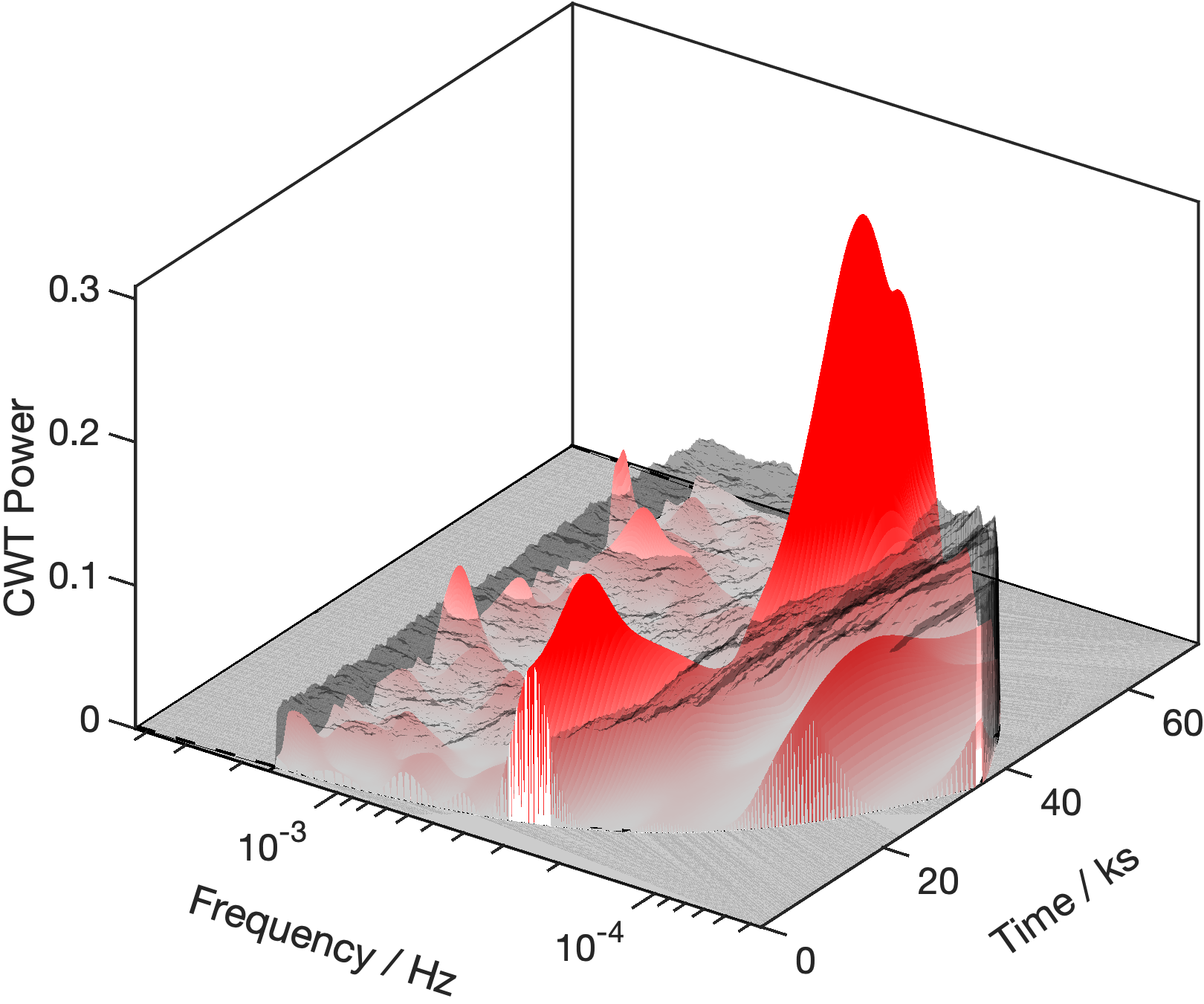}
    \caption{The wavelet power spectrum of a simulated time series consisting of an underlying red noise continuum, with a Lorentzian component at $2.5 \times 10^{-4}$ Hz. The generalized Morse wavelet parameters used were $\{\gamma,P\} = \{3,60\}$. Left: The shade of red represents the wavelet power. The curved lines in the upper-right panel defines the cone of influence (COI).  The shaded (grey) regions below the COI and above the Poisson noise line mark the frequencies that should be treated with caution. The regions in the dashed contour lines represent the amplitudes that are significant at the 90\% level ($S_{90}$), and the solid contour lines are are significant at the 99\% level ($S_{99}$). In the PSD on the left, the vertical line marks the Poisson noise level and the dotted line is the best-fit to the binned PSD used to estimate the underlying noise in the light curve (lower panel).  Right: The 3-dimensional representation of the left panel. The grey surface is the plane of significance at the 95\% level, i.e. $M_{95}$. All points of the wavelet power spectrum that are at a greater CWT power (i.e. wavelet power or amplitude) than $M_{95}$ are in the region $S_{95}$. For both panels, $S$ were calculated using 1000 simulated light curves, consisting of a simple power law PSD (without the Lorentzian component present). Each subsequent wavelet power spectrum plot in this work uses the same generalized Morse wavelet parameters, and respective confidence levels for the 2-dimensional and 3-dimensional representations. The `rms-squared' normalization \citep{uttley2014} was used for each PSD in this work.}
    \label{fig:red_noise_plus_qpo_cwt_2d_and_3d}
\end{figure*}

\subsection{Interpretation of the wavelet power spectrum}

One challenge of the wavelet analysis is that the wavelet power spectrum is a surface in 3-dimensions and might be more difficult to quantify than a Fourier transform.  Not all features in the image are necessarily important or significant.  Understanding how to interpret the results correctly, and identify artifacts that arise from either statistical fluctuations or from the nature of the wavelet transform itself, is paramount.

The wavelet power spectrum amplitudes are represented in  2-dimensions as a shade of red. In 3-dimensions these amplitudes are represented by both a height along an axis orthogonal to the $(t,f)$ plane, as well as the same shade of red used in the 2-dimensional representation. Due to long tails in the distribution of wavelet power spectrum amplitudes across the entire $(t,f)$ space and the visual limitations of using a colour to plot information, the entire range of wavelet power amplitudes are not represented equally. The ``maximum" colour was cut off at the $97.5^{\text{th}}$ percentile, so any amplitude at or greater than that percentile shows up as the darkest shade of red, and/or the highest height. Similarly, the ``minimum" percentile was cut off at the $0.1$ percentile.

It is important to note that the wavelet power spectrum amplitudes (i.e. the colour of red) are relative in each spectrum.  One wavelet power spectrum can not be directly compared to another because the colours and amplitudes will be different.  

\subsubsection{Cone of influence}

The finite length of the light curves implies that not all frequencies can be examined with significance at any given time.  From the onset of the observation, higher frequencies can be examined at earlier times than lower frequencies.  As the observation trails off, again the examination of lower frequencies will have to be concluded prior to study of the higher frequencies.  This produces a so-called cone of influence (COI) in the time--frequency space of the wavelet power spectrum, which corresponds to a region where the data can be best studied (Fig.~\ref{fig:red_noise_plus_qpo_cwt_2d_and_3d} or any other figure).

There is no precise mathematical definition for the COI, which depends on the choice of wavelet basis vector.\footnote{For detailed descriptions and definitions of the COI, see \cite{torrence1998}, \cite{nobach2007review}, and \cite{lilly2017element}.} In the case of the generalized Morse wavelet, the COI definition also depends on the parameters $P$ and $\gamma$.  In the {\sc matlab} {\sc wavelet toolbox} implementation of the generalized Morse wavelet with an input time series, the COI is defined as a function of frequency and calculated as \citep{misiti1996wavelet}:

\begin{align}
    \text{COI}(f) = \frac{f_c \sigma_{t,\psi}}{2\pi S dt}.
\end{align}

Here, $f_c$ is the central frequency of the generalized Morse wavelet, which is a function of $P$ and $\gamma$. The time series bin width is $dt$ and $S$ is a vector called "samples" which is a function of the time series length $N$. Each element of $S$ takes on an integer value, starting at $1$, stepping to $N/2$ at the halfway point, and then stepping back down to 1. For an even $N$, $S$ takes on the values: 
\begin{align}
    S = (1,2,...,N/2,N/2,...,2,1),
\end{align}
and for an odd $N$:
\begin{align}
    S = (1, 2, ...,(N + 1)/2, ..., 2, 1).
\end{align}

The wavelet standard deviation is defined in \citet{lilly2008_morse_wavelets} as:

\begin{align}
    \sigma_{t,\psi} = f_c \frac{\int |\Psi'(f)|^2 df}{\int |\Psi(f)|^2 df}
\end{align}
where $\Psi(f)$ is the wavelet basis vector projected to the frequency domain and $\Psi'(f) = d\Psi/df$. For the generalized Morse wavelet, this is:

\begin{align}
    \Psi_{P,\gamma} (f) = a_{P,\gamma} f^{\frac{P^2}{\gamma}} e^{- \gamma f}.
\end{align}

When plotting the COI of the wavelet power spectrum as an image projected to 2-dimensional time--frequency space, the regions below the COI are shaded as a reminder to treat any apparent wavelet peaks in that region with skepticism, as they could be due to the boundary effects which arise from the finite time series length (e.g. Fig.~\ref{fig:sin_waves_cwt_for_coi_eg}). The frequency at which the Poisson noise begins to dominate is calculated using the Poisson noise power as defined in \citet{uttley2014}.  The region above the Poisson noise level is also shaded to indicate that it is dominated by noise. 

When plotting the wavelet power spectrum as a surface in 3-dimensional time--frequency--amplitude space, the regions below the COI and in the Poisson noise are fully subtracted to zero (e.g. Fig.~\ref{fig:red_noise_plus_qpo_cwt_2d_and_3d}). This is simply done to improve clarity.

\subsubsection{Confidence intervals} \label{significance_section}

An important aspect of the wavelet power spectrum, is to determine intervals in the  time--frequency space where the amplitudes of the wavelet power spectrum are significant. In \citet{torrence1998}, a method is outlined to estimate the significant regions, for Gaussian distributed time series which follow a red-noise distribution. Since AGN light curves typically have log-normal, or generally non-Gaussian flux distributions, a Monte Carlo process is adopted in this work to estimate the significant regions.

The significant regions $S_{\alpha}(t,f) \in W(t,f)$, where $\alpha$ is the confidence level, were calculated using light curve simulations described in Section \ref{lc_sim_section}.  First, an ensemble of light curves $x_{ensemble}(t)$ mimicking each observation were simulated. Here, each ensemble consisted of 1000 light curves. The input parameters for these simulations were: the observed flux distribution; the PSD slope based on a simple power law fit to the observed binned PSD; the number of data points; and the light curve duration.  From this information, an ensemble of wavelet power spectra $W_{ensemble}(t,f)$ were calculated from the simulated light curves. From $W_{ensemble}(t,f)$ and a desired confidence level $\alpha$, the $\alpha^{\text{th}}$ percentile can be calculated at each $(t_i,f_i)$. This set of points forms a ``plane of significance'' $M_{\alpha}(t,f)$ in the 3-dimensional plot (right panel of Fig.~\ref{fig:red_noise_plus_qpo_cwt_2d_and_3d}), and the significant region $S_{\alpha}$ consists of all the points $W > M_{\alpha}$.  The derived confidence contours and intervals are shown in Fig.~\ref{fig:red_noise_plus_qpo_cwt_2d_and_3d} for a simulated time series.

\subsection{Simulation tests} \label{sim_section}

A variety of simulations are carried out to determine the effectiveness of distinguishing and identifying (periodic) signals in AGN light curves.  Here we examine light curves with different power distributions, periodic signals, and with non-stationary behaviour to better understand the appearance of the wavelet power spectrum.

\subsubsection{Different colours of noise} \label{colours_of_noise_section}

Light curves with power spectra in the form of $f^{-\beta}$ are examined to visualize the effect of different values of $\beta$ on the wavelet image.  AGN commonly have $\beta \sim 2$ (red noise), but for simulation purposes $\beta$ ranged from 0 (white noise) to 3 (black noise).

\begin{figure}
    \centering
    \includegraphics[width=\columnwidth]{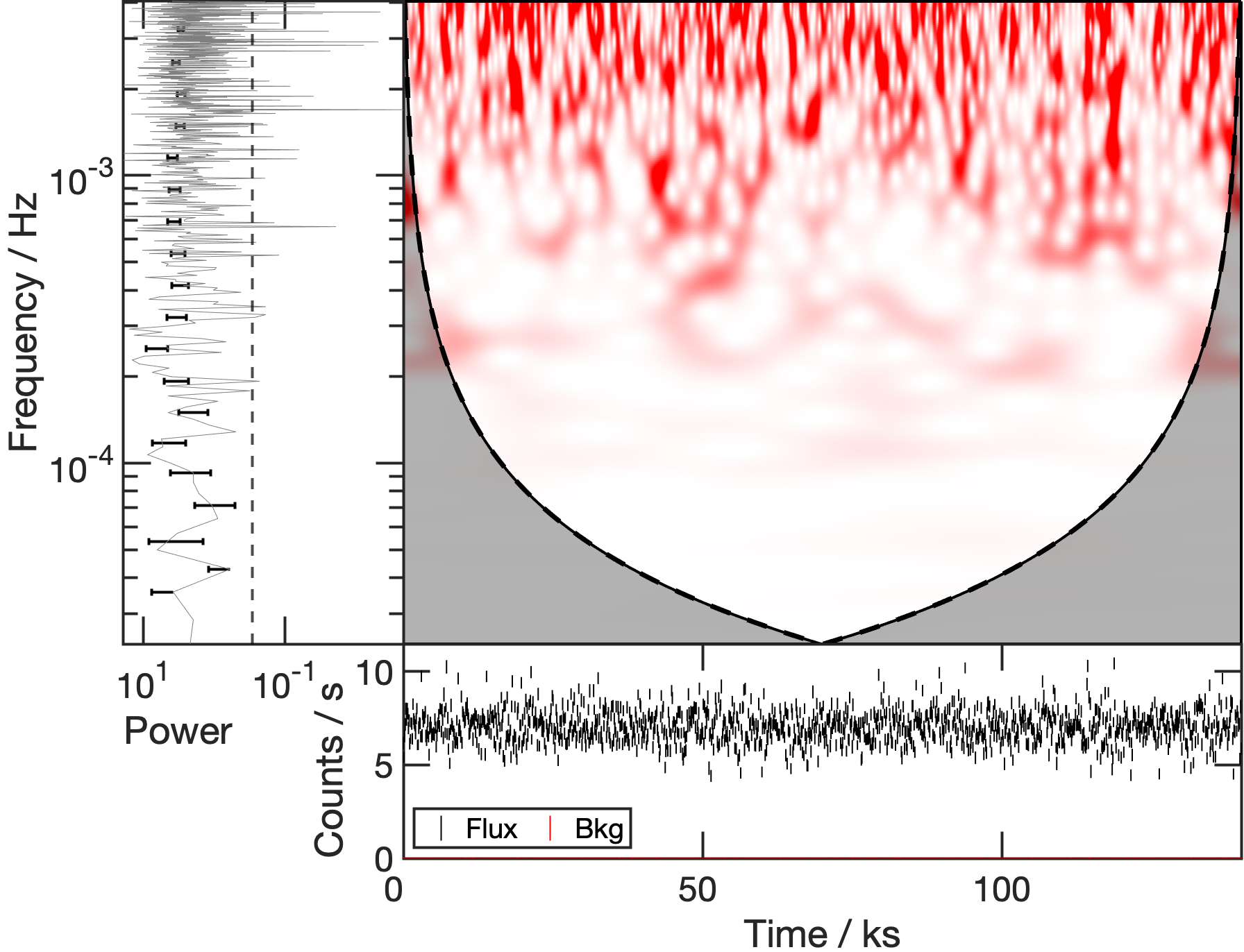}\\
    \includegraphics[width=\columnwidth]{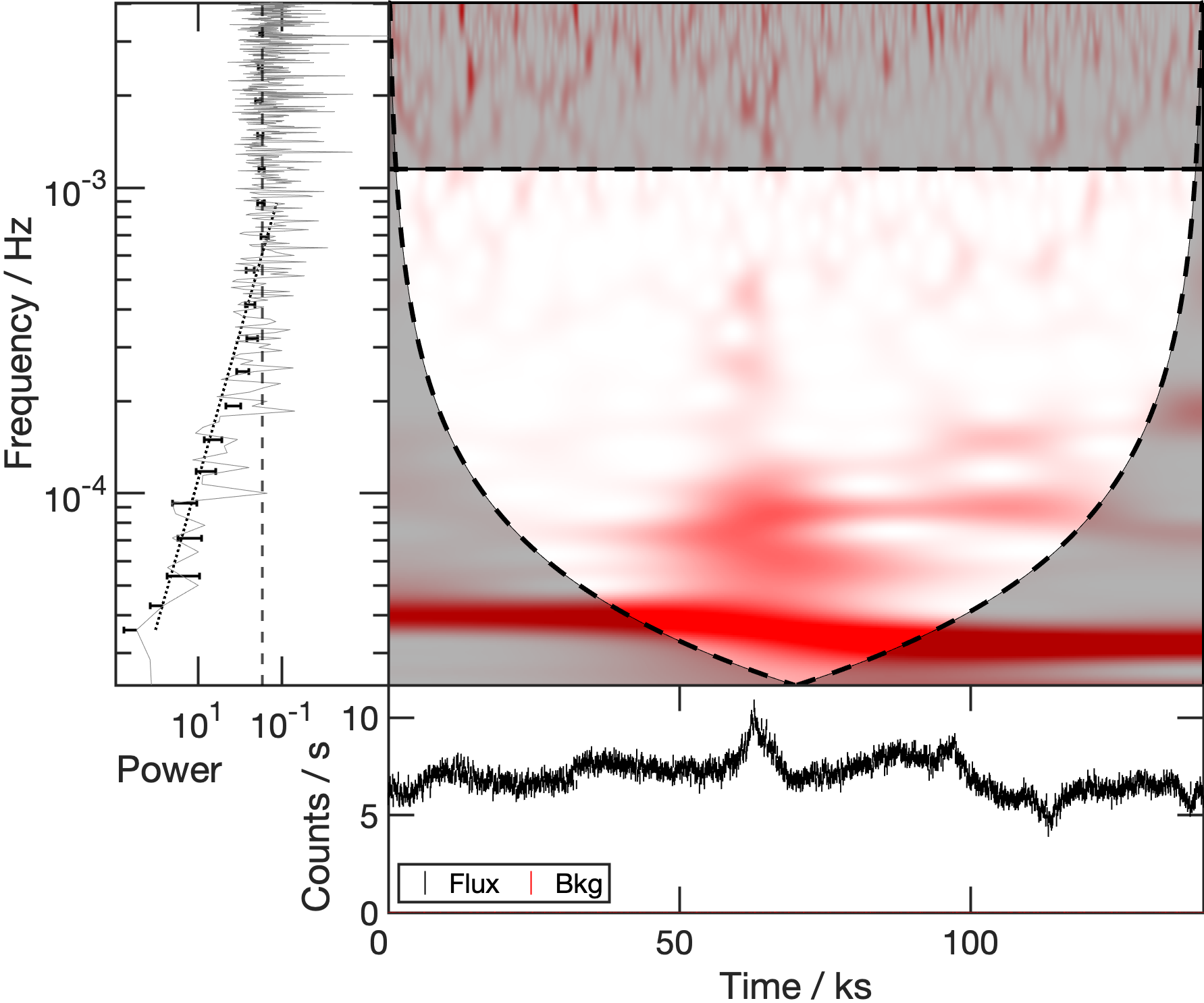}
    \caption{Wavelet power spectra of two simulated time series consisting of pure power law noise. As the time series are pure noise, no confidence levels are plotted. The light curves have input $\beta=0$ (top panel) and $\beta= 3$ (lower panel) PSDs. An example with $\beta=2$ was already shown in Fig.~\ref{fig:red_noise_plus_qpo_cwt_2d_and_3d}.}
    \label{fig:noise_cwts}
\end{figure}

A simulation with $\beta=2$ is already shown in Fig.~\ref{fig:red_noise_plus_qpo_cwt_2d_and_3d}.  In Fig.~\ref{fig:noise_cwts}, examples with $\beta=0$ and $\beta=3$ are displayed.  There are no true (periodic) signals in either of these tests.

The random noise in the wavelet power does depend on index $\beta$.  At $\beta=0$, much of the random power is evident at high frequencies.  As $\beta$ is increased, the wavelet power amplitudes are skewed towards lower frequencies where the PSD has more power.  This behaviour is largely expected, but highlights that signal detection may depend on the underlying power spectrum shape.

\subsubsection{Different types of signals}

\begin{figure}
    \centering
    \includegraphics[width=\columnwidth]{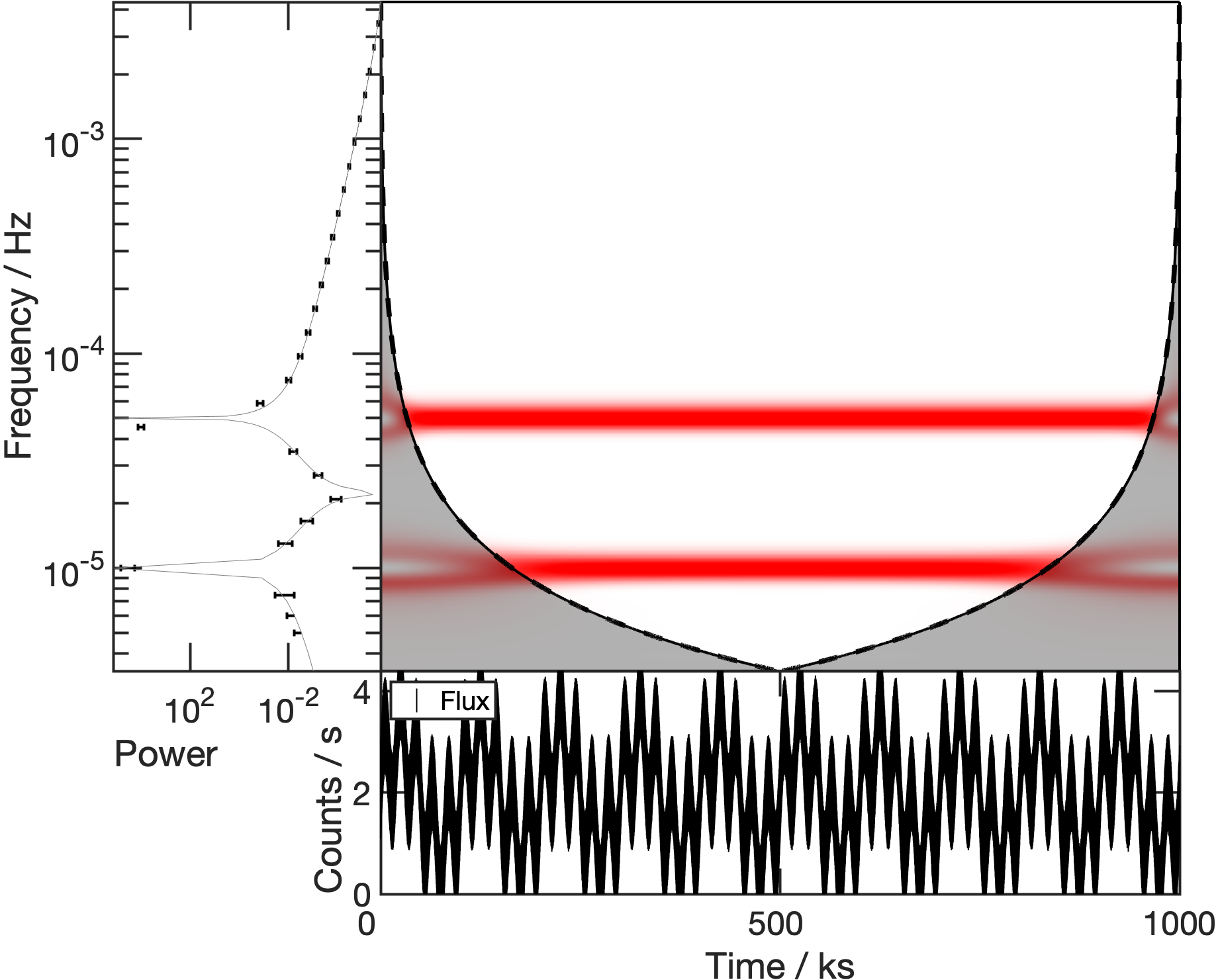}
    \includegraphics[width=\columnwidth]{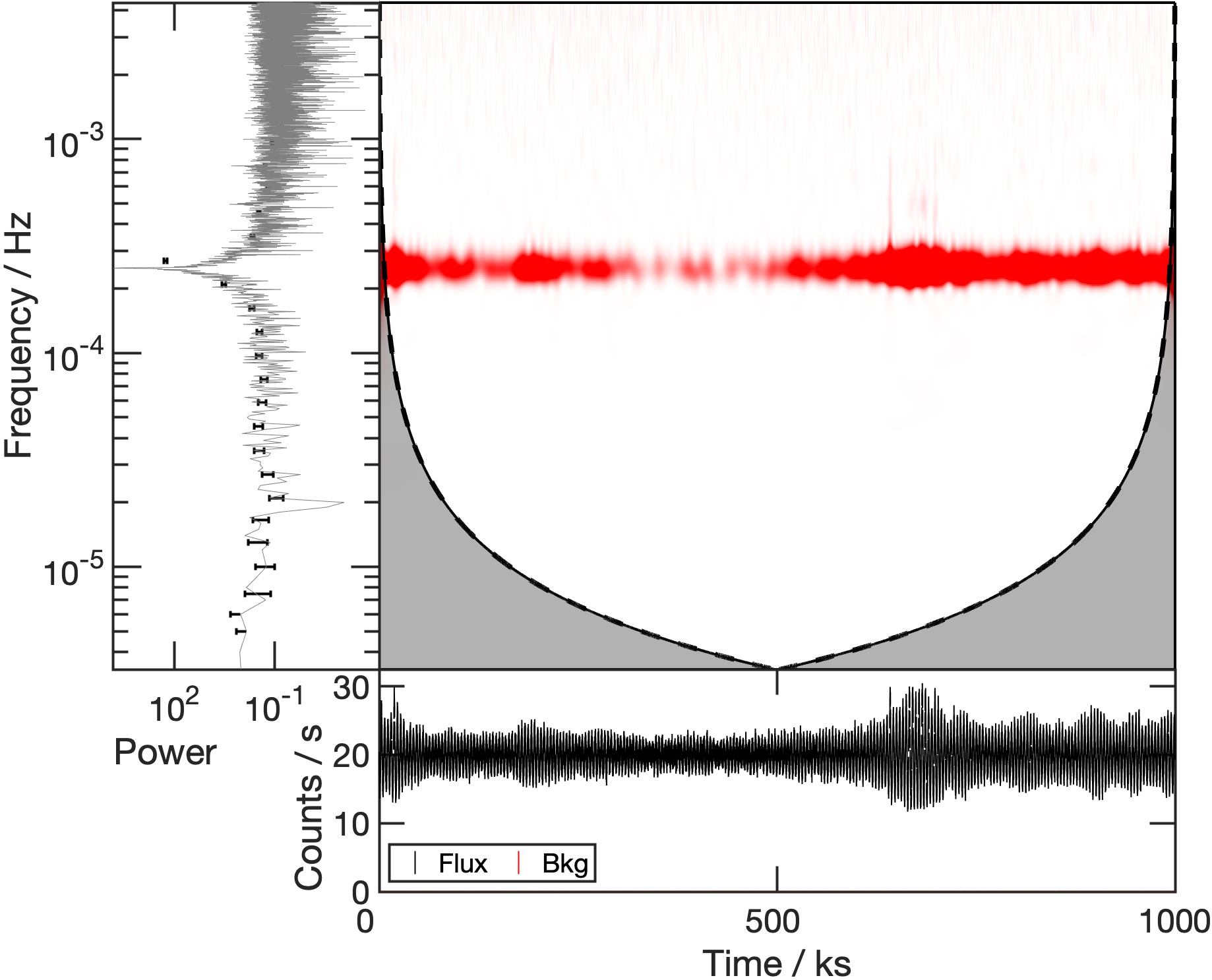}
    \caption{Top panel: The wavelet power spectrum of the function $f(t) = \sin(2\pi f_1 t) + \sin(2\pi f_2 t)$, where $f_1 = 10^{-5}$ Hz and $f_2 = 5 \times 10^{-5}$ Hz. Lower panel: A simulation of a time series consisting of a single Lorentzian profile at $2.5 \times 10^{-4}$ Hz and with a full-width half-maximum of $2 \times 10^{-5}$ Hz.}
    \label{fig:sin_waves_cwt_for_coi_eg}
\end{figure}

The most basic pure component in a signal would be a perfect sine wave. The wavelet power spectrum of two sine waves is shown in Fig. \ref{fig:sin_waves_cwt_for_coi_eg} (top panel). Here it can be seen that since there is no variability in the frequencies present in the signal, a pure sine wave results in a constant band in the wavelet power spectrum.

In the lower panel of Fig. \ref{fig:sin_waves_cwt_for_coi_eg}, the wavelet power spectrum for a single  Lorentzian component is shown.  Here it becomes evident that even though the signal is a persistent and single Lorentzian component, there is still variability in the overall amplitude of the signal since the Lorentzian does have some ``width'' to it.  Based on the colours in the wavelet spectrum, the feature appears to fluctuate, despite the fact that it is persistent in the input PSD. Therefore, due to the random variability intrinsic to the Lorentzian profile, a persistent peak in the PSD may not be always appear constant with time in the wavelet power spectrum.

\subsubsection{The effects of count rate and exposure for detecting periodic signals}

\begin{figure}
    \centering
    \includegraphics[width=\columnwidth]{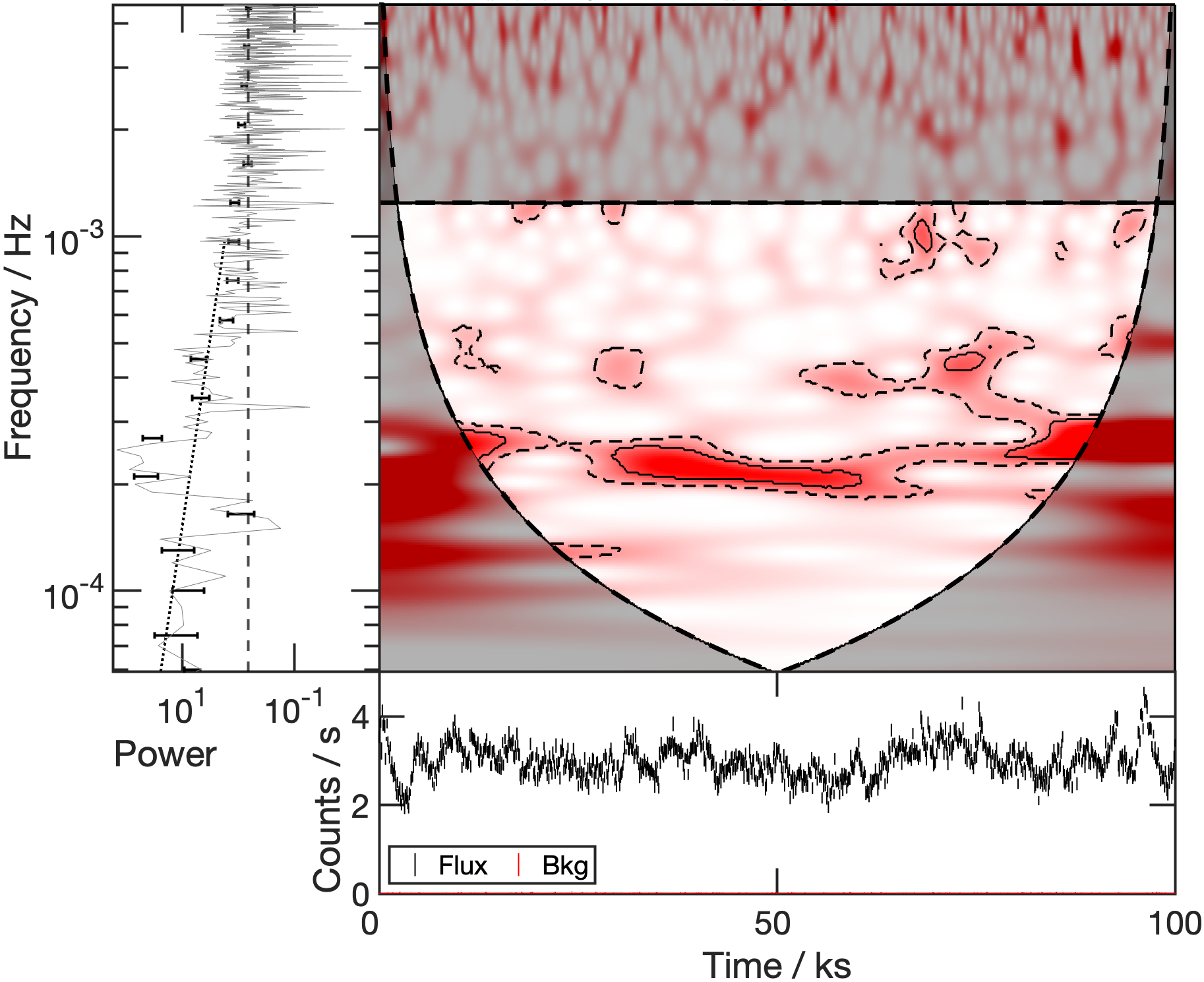}\\
    \includegraphics[width=\columnwidth]{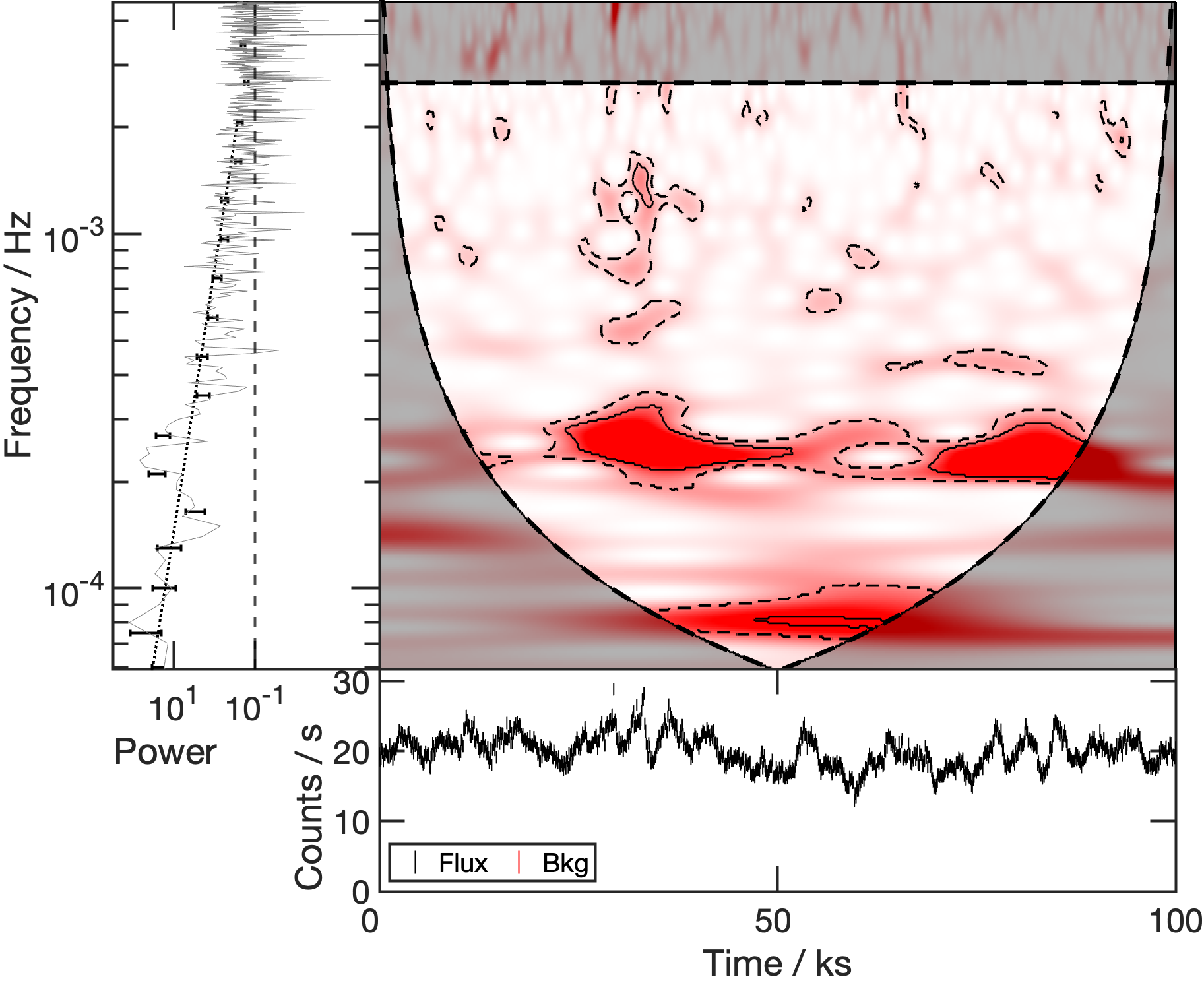}
    \caption{An example of two simulated time series consisting of identical input PSDs of a power law with $\beta = 1.7$ and a Lorentzian profile centered at $2.5 \times 10^{-4}$ Hz. These differ only by average count rate (3 on top and 20 below).}
    \label{fig:cwt_changing_avg_count_rate}
\end{figure}

Adjusting the average count rate by multiplying the time series by a constant will not change the relative wavelet amplitudes throughout the $(t,f)$ space. The absolute amplitudes of both $W$ and $M$ will be affected, but this has no effect on the ability to detect a significant region.  This is demonstrated in the wavelet power spectra in Fig. \ref{fig:cwt_changing_avg_count_rate} showing that the overall scaling will not affect which ($t,f)$ regions are significant, since both $W$ and $M$ scale identically. From this, we see that wavelet timing methods can be useful even for dim objects.

\begin{figure}
    \centering
    \includegraphics[width=\columnwidth]{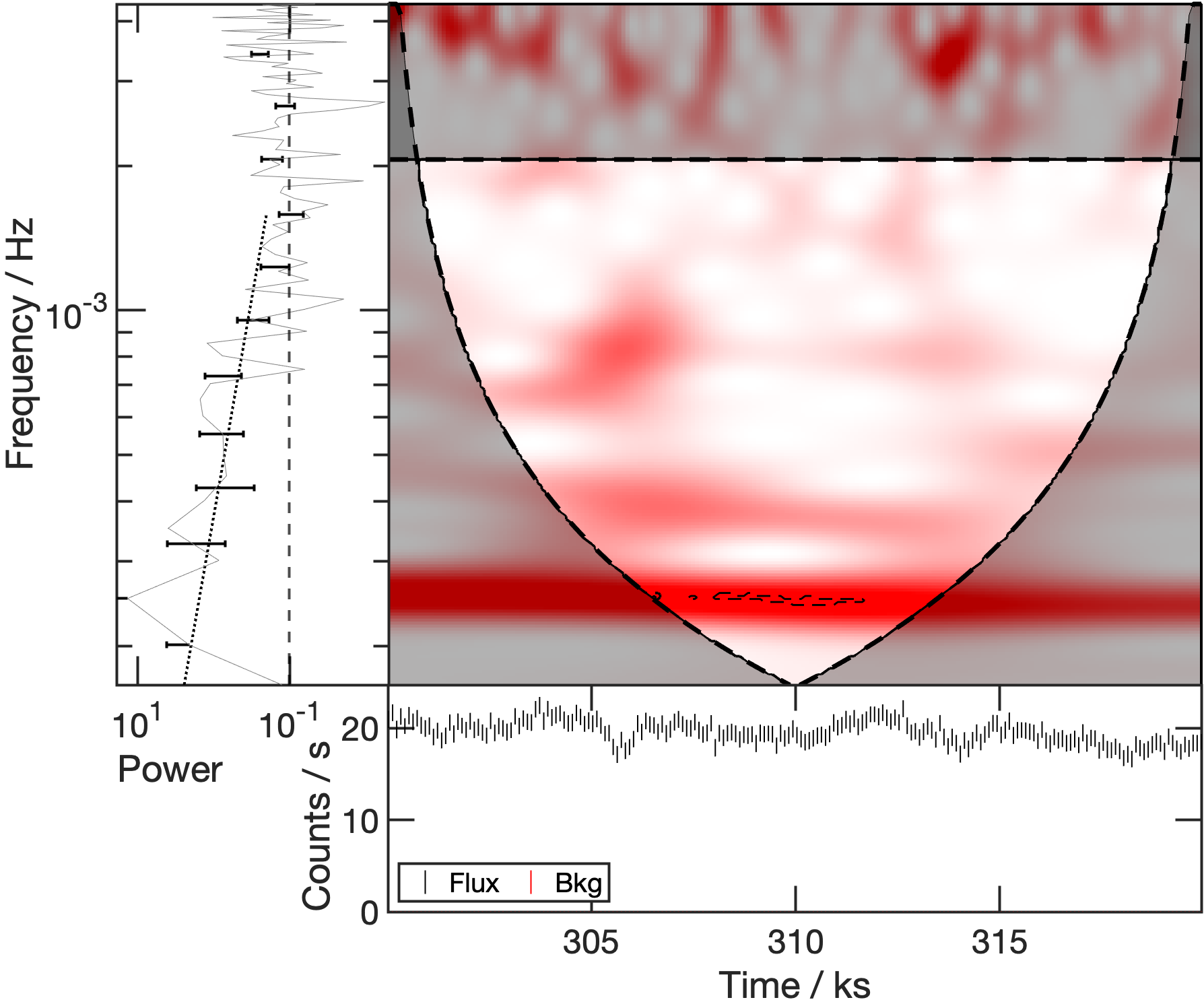}\\
    \includegraphics[width=\columnwidth]{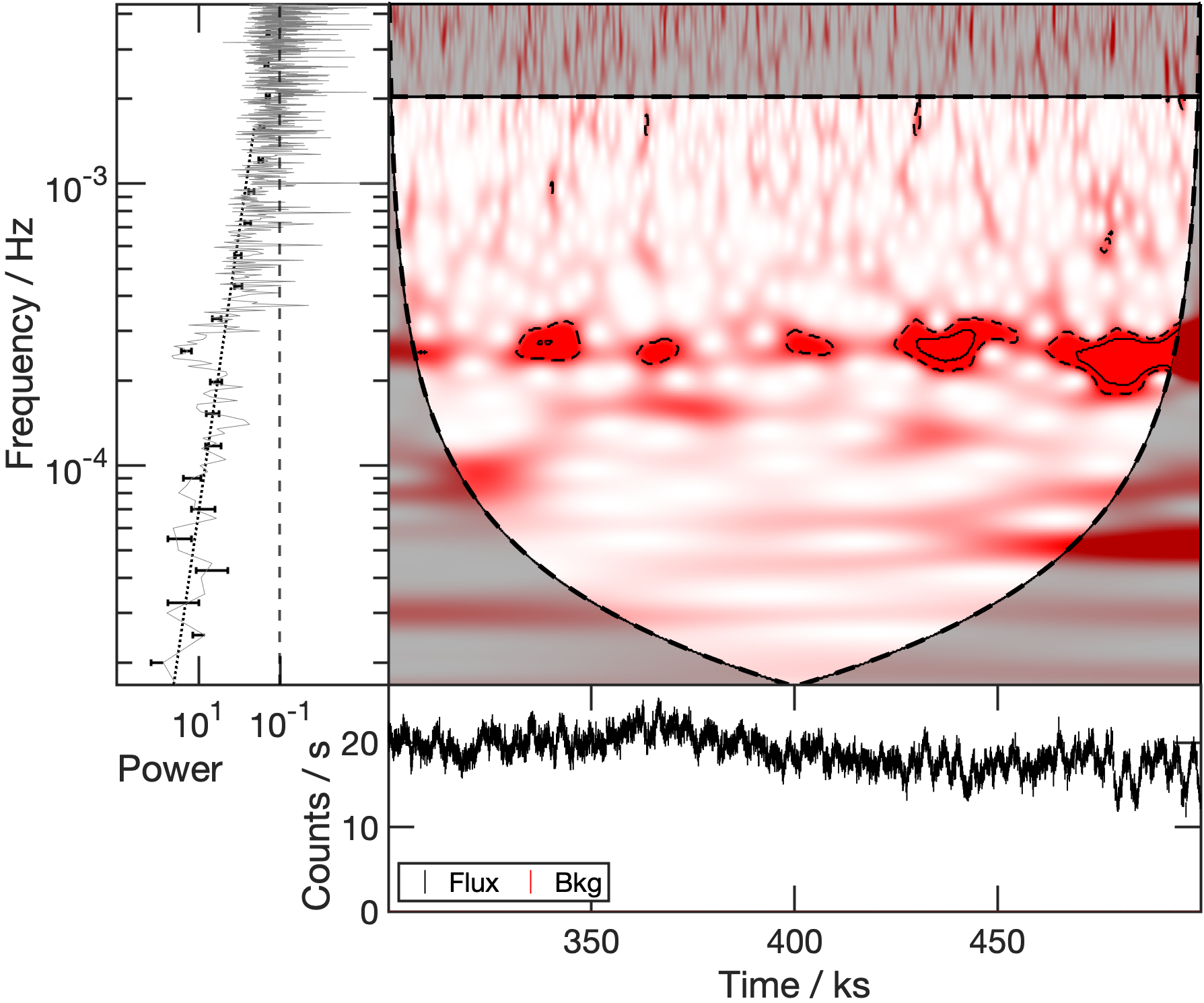}
    \caption{An example of two simulated time series that are each a subset from the same 1 Ms time series, consisting of a power law with $\beta = 1.7$ and a Lorentzian profile centered at $2.5 \times 10^{-4}$ Hz. The top panel is 20 ks long, and the bottom panel is 200 ks long.}
    \label{fig:cwt_changing_exposure}
\end{figure}

Another important aspect in X-ray observations is the exposure time needed to detect a feature in the light curve. This is shown in Fig. \ref{fig:cwt_changing_exposure}, where the wavelet power spectrum is calculated for a 20 ks and 200 ks light curve drawn from a long 1 Ms simulation.  Here, we can see that a shorter exposure will limit the frequency range that can be examined significantly.  As with the Fourier transform, repeated occurrences of a signal will improve the potential of detecting it with significance.  With longer exposures, there is a stronger possibility of the signal being significant in the wavelet power spectrum.

\subsubsection{The effects of the PSD and signal strength}

\begin{figure}
    \centering
    \includegraphics[width=\columnwidth]{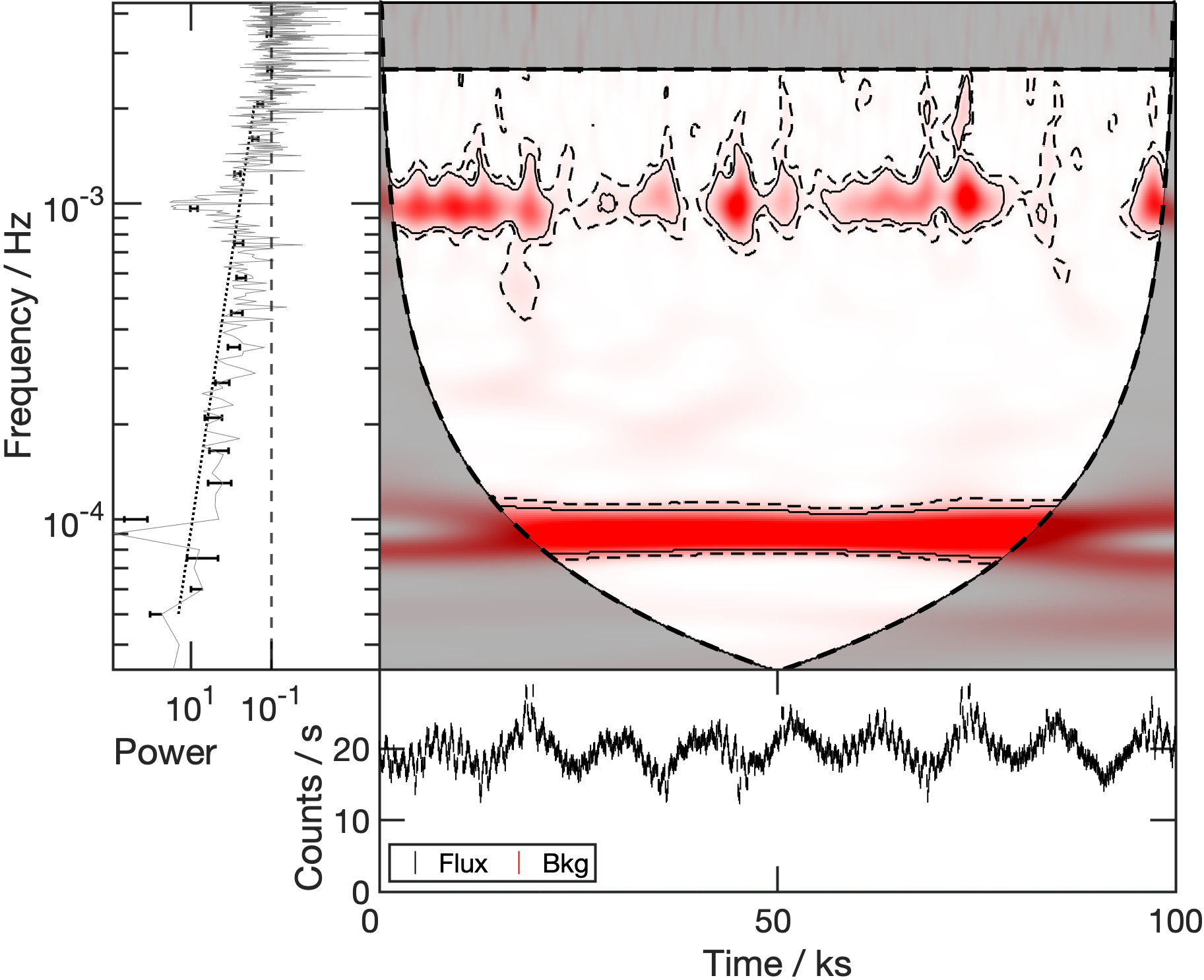}\\
    \includegraphics[width=\columnwidth]{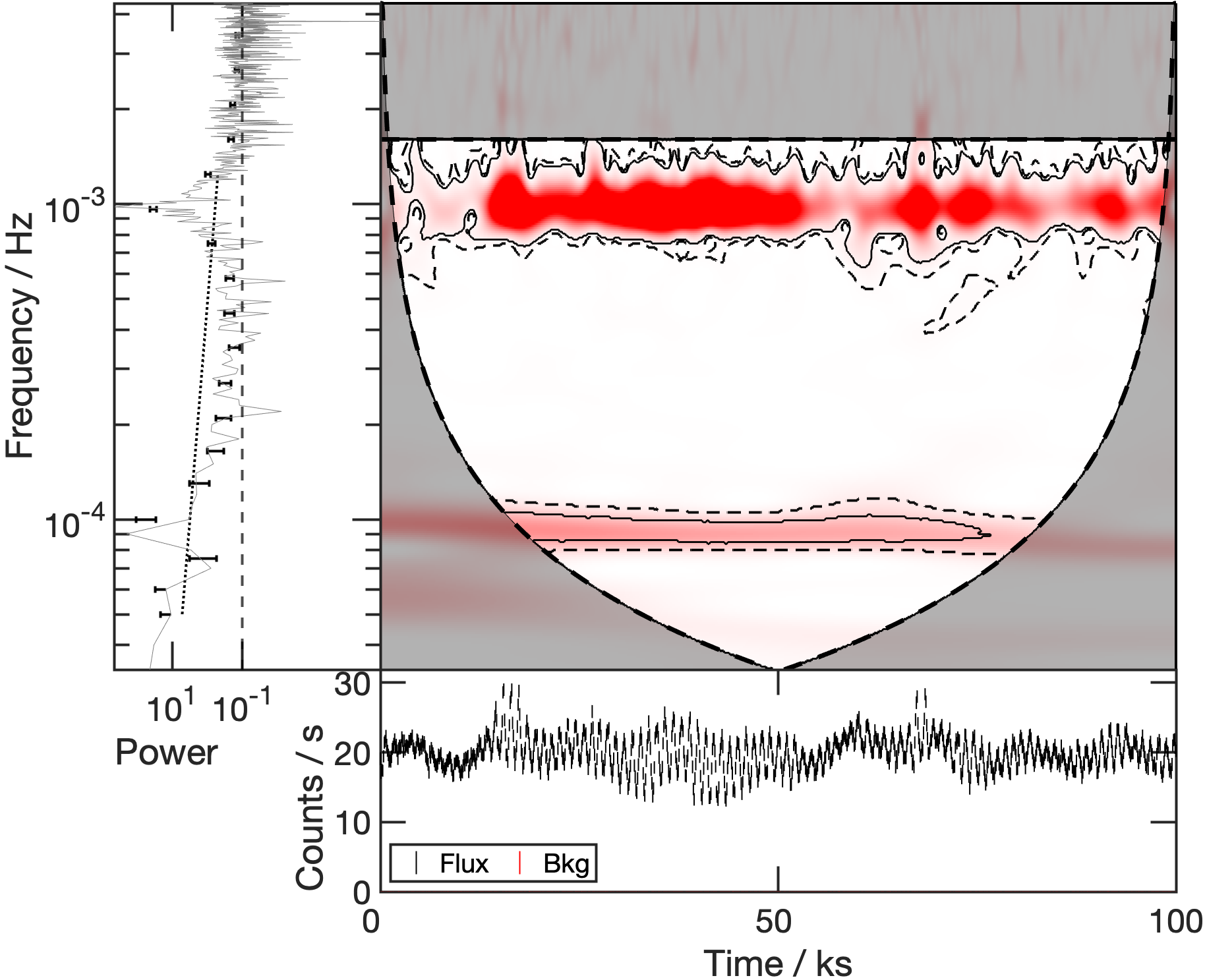}
    \caption{Wavelet power spectra of two simulated time series consisting of an underlying power law and two Lorentzian profiles centered at $10^{-4}$ Hz and $10^{-3}$ Hz. Top panel: The power law is $\beta = 2$, and the lower frequency Lorentzian appears to be more significant in the wavelet power spectrum. Bottom panel: The power law is $\beta = 3$, and the higher frequency Lorentzian appears to be more significant in the wavelet power spectrum.}
    \label{fig:cwt_2qpo_changing_pl}
\end{figure}

The input power spectra from the previous section are modified by including a Lorentzian profile at a given frequency to replicate a periodic signal.  In Fig.~\ref{fig:sin_waves_cwt_for_coi_eg}, it is clearly demonstrated that a periodic signal is straightforward to detect with a wavelet power spectrum in the absence of noise.  Fig. \ref{fig:cwt_2qpo_changing_pl} shows that the underlying power law will affect which regions are significant in the wavelet power spectrum.

\begin{figure}
    \centering
    \includegraphics[width=\columnwidth]{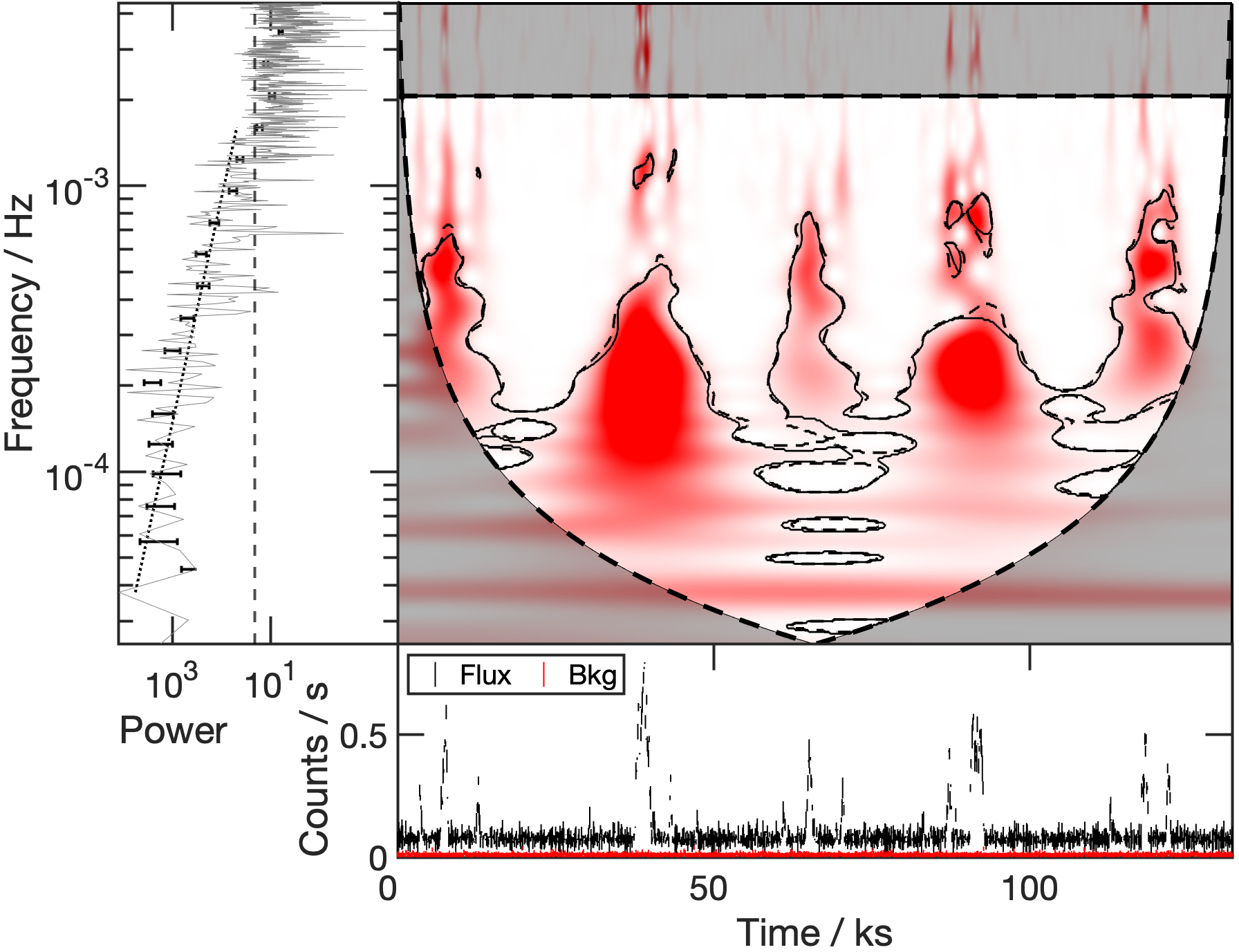}
    \caption{A simulated time series with a white noise PSD and consisting of ten Lorentzian components spaced in frequency and power. }
    \label{fig:simulated_qpe}
\end{figure}

In addition, we also examine the potential of identifying multiple periodic signals.  For this, multiple Lorentzian profiles were added to PSDs with different slopes.   Again, in the absence of noise, Fig.~\ref{fig:sin_waves_cwt_for_coi_eg} demonstrates that multiple signals are easily distinguished.  In Fig.~\ref{fig:simulated_qpe}, we examine an extreme case where ten Lorentzian profiles are added to an underlying white noise spectrum.  The Lorentzian profiles are spaced in frequency to resemble harmonics of equal power. The frequencies are spaced by a geometric factor of 3/2, which is the harmonic spacing of QPOs in stellar mass X-ray binaries. The lowest frequency harmonic is $(9 \text{ hr})^{-1} = 3 \times 10^{-5}$ Hz, which is the lowest harmonic present in the AGN GSN 069 \citep{miniutti2019nine_gsn}. For this simulation, the harmonics all had equal power. Multiple features stand out as potentially interesting. In such a situation, it is challenging to accurately model the underlying power law, as can be seen in fitted PSD in Fig.~\ref{fig:simulated_qpe}.  With all the signals present and overlapping, the best-fit power law has $\beta = 1.26$, even though the input was $\beta=0$.  However, the wavelet spectrum clear indicates that multiple frequencies might be important.

\begin{figure}
    \centering
    \includegraphics[width=\columnwidth]{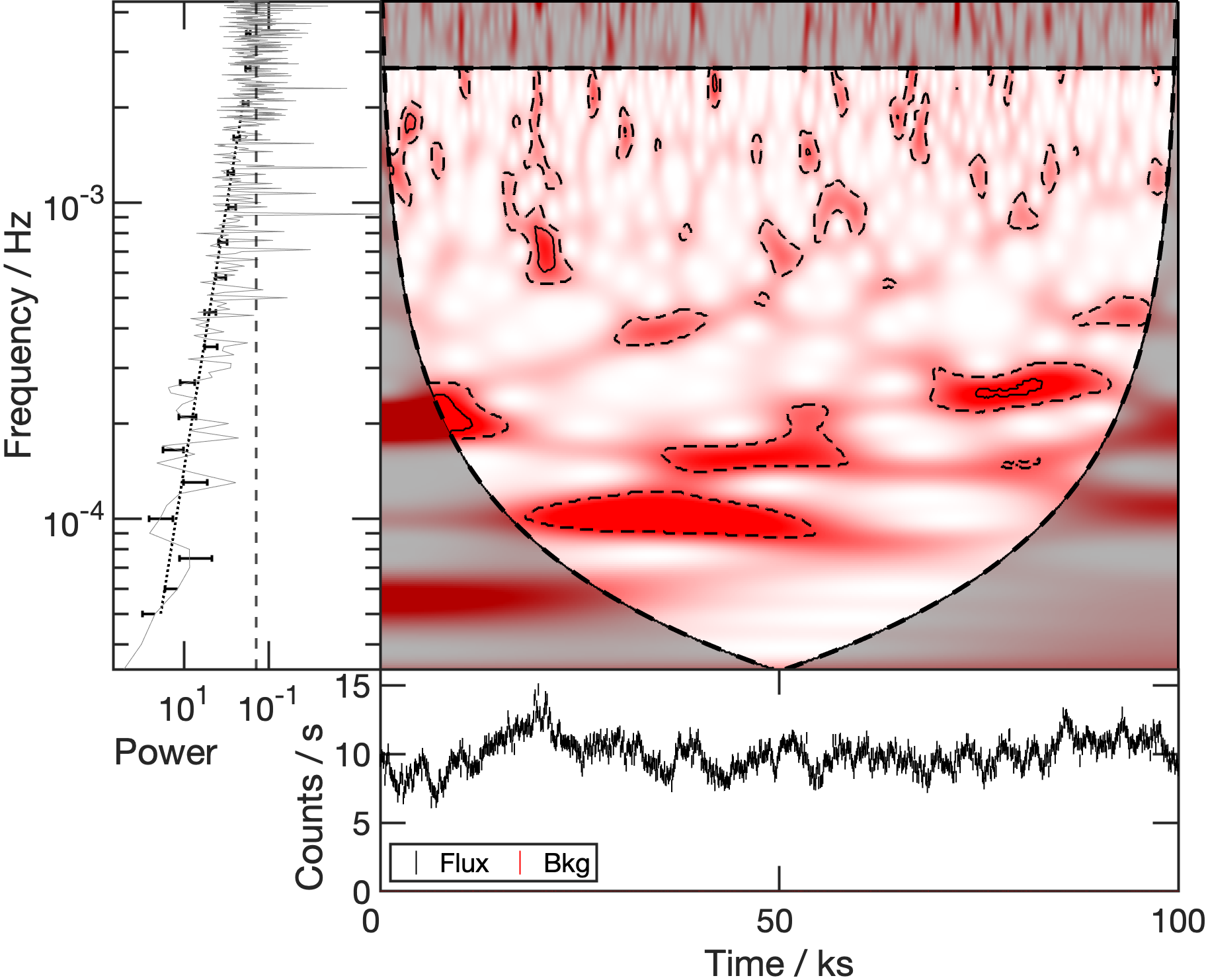}//
    \includegraphics[width=\columnwidth]{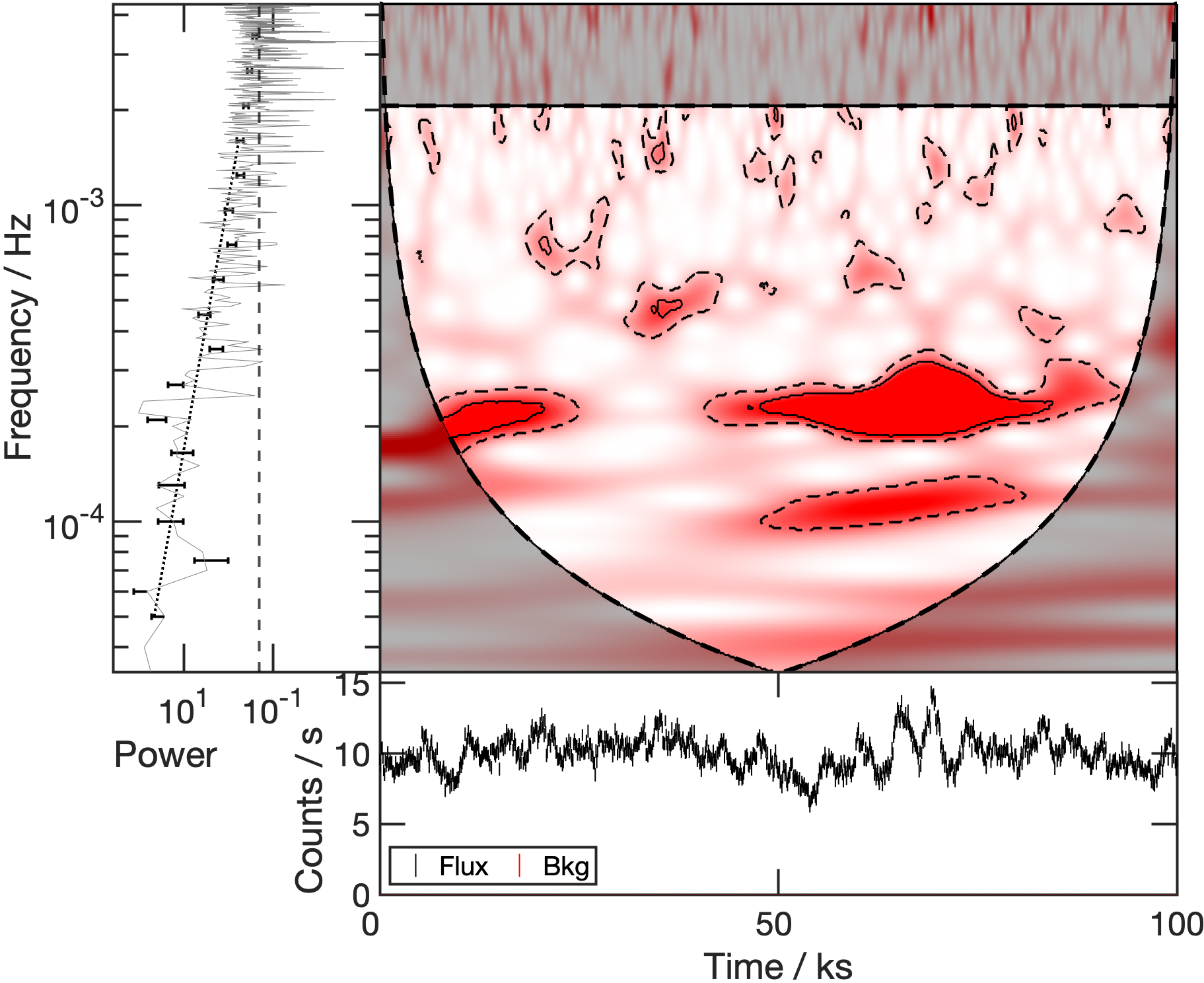}//
    \includegraphics[width=\columnwidth]{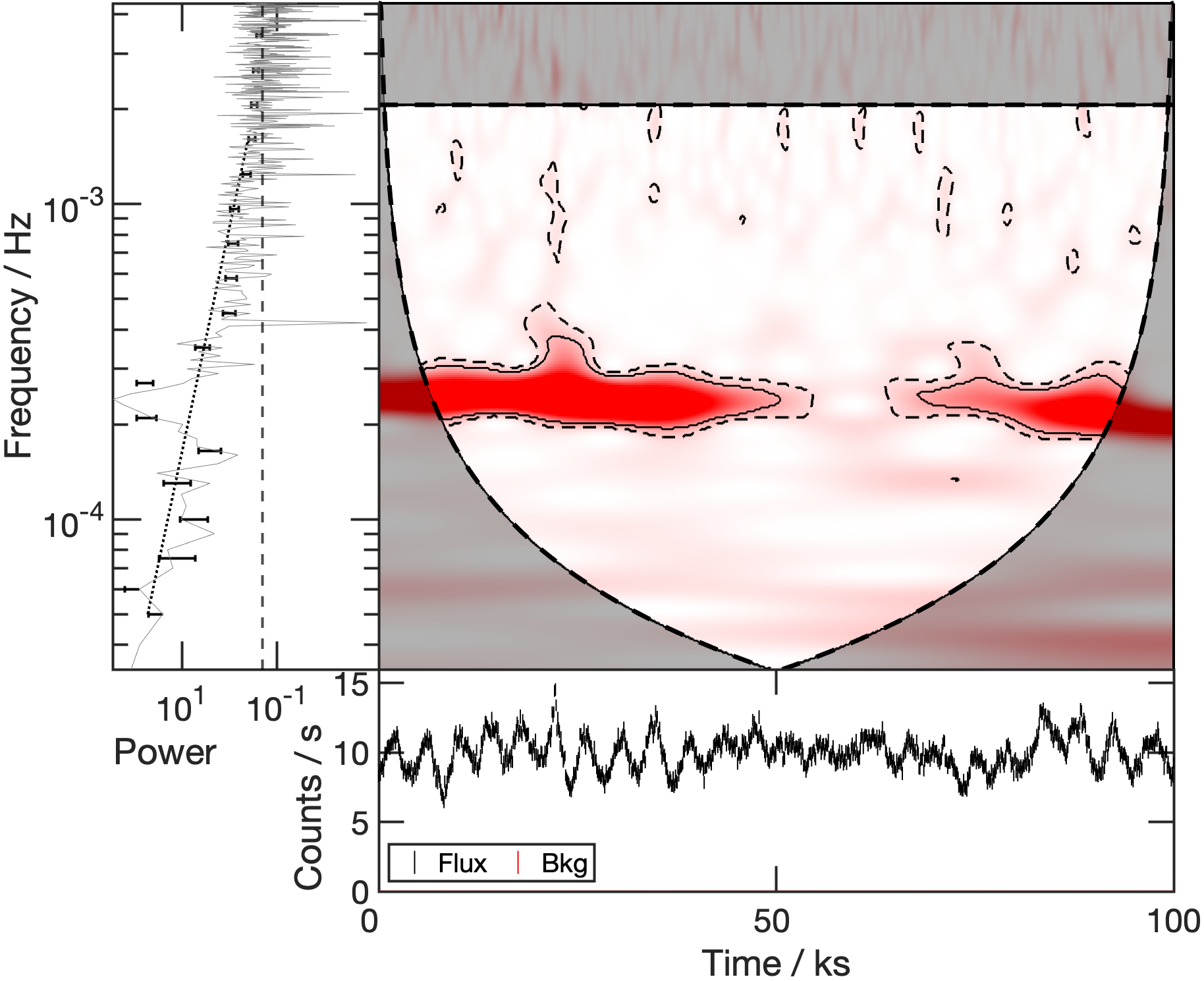}
    \caption{Simulated time series consisting of a Lorentzian component centered at $2.5 \times 10^{-4}$ Hz, and a power law component with $\beta = 1.7$. The ratio of the Lorentzian/power law respective integrals are 1\% (top panel), 10\% (middle panel), and 100\% (bottom panel). }
    \label{fig:different_L_PL_ratios}
\end{figure}

We also examined the effect on the wavelet power spectrum of varying the ratio of power between the Lorentzian component and power law slope, by dividing the integral of the Lorentzian component by the integral of the power law component. In Fig. \ref{fig:different_L_PL_ratios} it is apparent that when the Lorentzian component is very weak compared to the power law component, for example around 1\% of the strength, it does not appear to be significant in the wavelet power spectrum despite being persistent. At 10\% the Lorentzian component is significant but not fully persistent in the wavelet power spectrum. At 100\% the Lorentzian component is significant, and fully persistent other than a small segment in the middle of the time series, due to the random variance of the Lorentzian strength as shown in Fig. \ref{fig:sin_waves_cwt_for_coi_eg}. Note that for the \textit{XMM-Newton} observation of RE J1034+396 (Obs ID: 0506440101), the Lorentzian to power law ratio was approximately 10\%, and was fully persistent at the 90\% confidence level in the wavelet power spectrum (see Section \ref{rej1034396}).

\subsubsection{Non-stationary signals}

Non-stationarity was simulated by approximating the non-stationary time series as an ensemble of stationary light curves, i.e. a non-stationary piecewise PSD as suggested in \citet{alston2019non_stationary}. Different tests of non-stationary time series were completed, to see how the wavelet power spectrum coefficients change with time. Tests included:
\begin{itemize}
    \item varying the overall PSD power
    \item varying the overall PSD power with a constant power and frequency Lorentzian
    \item varying the PSD slope $\beta$
    \item varying the Lorentzian power with constant frequency
    \item varying the Lorentzian frequency with constant power
    \item varying the PSD slope as a random flicker
    \item varying the PSD slope as a randomly increasing flicker
\end{itemize}

\begin{figure}
    \centering
    \includegraphics[width=\columnwidth]{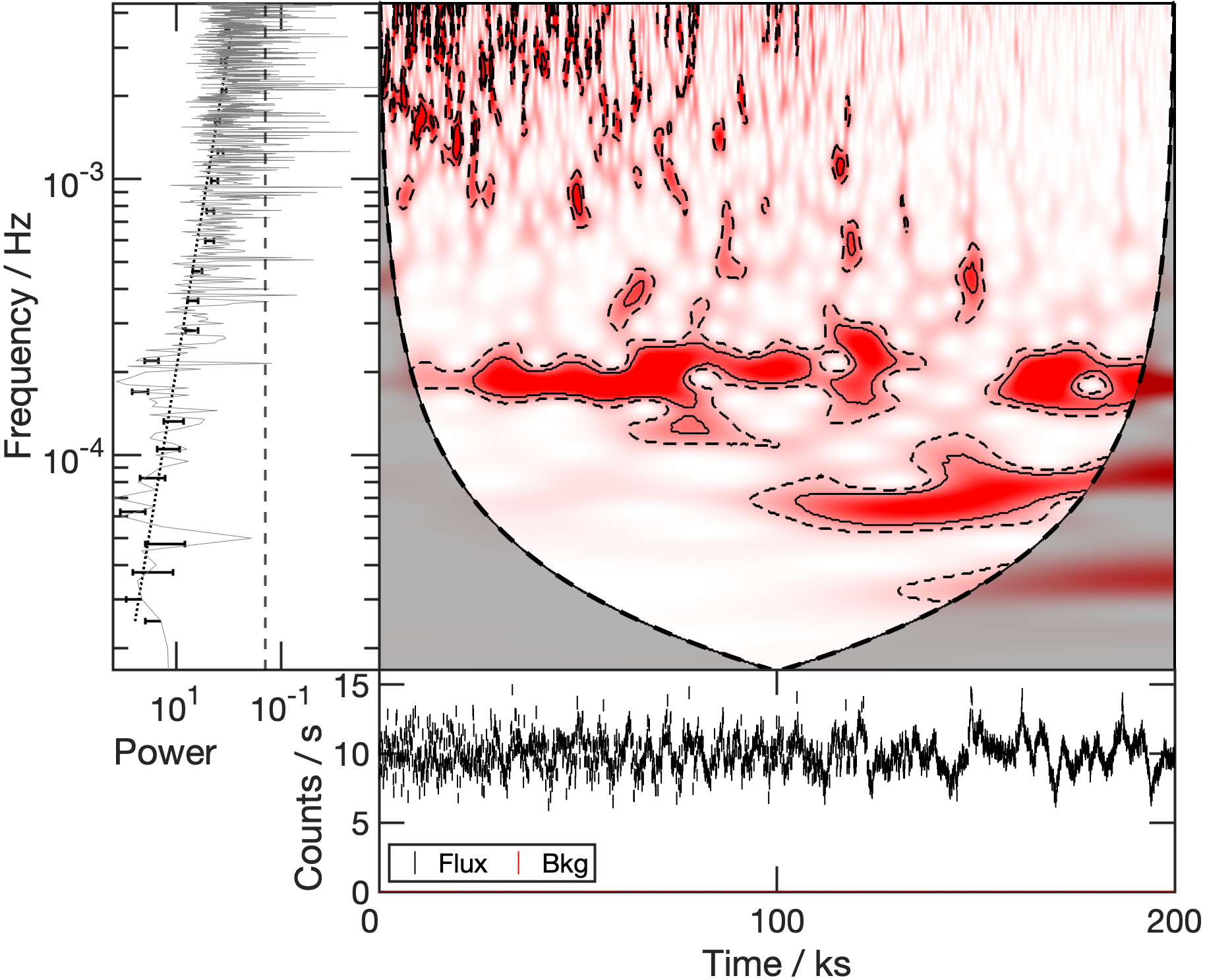}
    \caption{Simulation of a non-stationary time series, using a non-stationary piecewise PSD. The Lorentzian component is constant at $2.5 \times 10^{-4}$ Hz throughout the entire light curve. The power law component varies from $\beta=0$ to 3 linearly in 20 ks intervals. We can see that the Lorentzian component is still significant in the wavelet power spectrum, for different $\beta$ values. We can also see a general shift in the noisy amplitudes in the wavelet power spectrum, from higher to lower frequencies as the power law becomes steeper.}
    \label{fig:simulated_nonst_pl_const_qpo}
\end{figure}

In Fig.~\ref{fig:simulated_nonst_pl_const_qpo}, we show a test where we have a constant Lorentzian signal and an underlying PSD that is varying in slope.  The power law slope increases from $\beta=0$ to 3 linearly in 20 ks intervals while the Lorentzian profile remains constant at $2.5\times10^{-4}$ Hz. The degree of variability is likely extreme compared to real sources.

Fitting the averaged PSD with a single power law is obviously incorrect but still allows us to examine the wavelet spectrum for significant features.  A roughly persistent signal is present at the correct frequency, indicating that signals can examined even if the true underlying PSD slope is not well known. 

\section{Applications of wavelets to AGN and QPE X-ray light curves}

As a proof of concept, four AGN are used to examine the effectiveness of the wavelet analysis.  All of the objects have been observed with \textit{XMM-Newton} in the $0.3-10$~keV band, and their variability has been analysed with Fourier techniques or other methods.

We use archival data obtained from the \textit{XMM-Newton} \citep{Jansen+2001} Science Archive (XSA) for these four AGN, one of which exhibits QPE behaviour, to examine the effectiveness of the wavelet analysis on a wide range of variability behaviour. We focus on data collected by the EPIC pn \citep{Struder+2001} detector, which were processed in the following way using the \textit{XMM-Newton} Science Analysis System (SAS) version 20.0.0. From the \textit{XMM-Newton} Observation Data Files (ODFs) of each observation (see following subsections for specific observation IDs), we created event lists with the \textsc{epproc} task. These event lists were filtered using the \textsc{evselect} task with conditions $\textsc{pattern} \leq 4$ and $\textsc{flag} = 0$ to extract only single and double events. Source counts were extracted from a circular region of radius $35''$ centred on the source position (obtained from the NASA/IPAC Extragalactic Database), and background counts were extracted from a nearby off-source circular region of radius $50''$. Light curves were then created using the \textsc{epiclccorr} task and were rebinned to 100 s. We note that mild to negligible background flaring is present in each observation, however, we chose not to filter out these segments of the light curves in order to preserve light curve continuity. The effect of background flaring in the wavelet analysis is discussed, where applicable. Finally, we note that only RE J1034+396 exhibited significant photon pile-up (evaluated using the \textsc{epatplot} task), with $\sim10~\mathrm{per~cent}$ in single and $\sim30~\mathrm{per~cent}$ in double events over the $0.5-2~\mathrm{keV}$ band; all other sources had $<5~\mathrm{per~cent}$ and $<10~\mathrm{per~cent}$ pile-up in single and double events, respectively. Since here we investigate only broad band light curves such levels of pile-up should not significantly impact the interpretation of our results, and therefore we perform all forthcoming analysis on the uncorrected light curves.

Since we are outlining a preliminary method for using wavelet transforms, in general, the objects selected exhibit distinct behaviour from each other.  Here, we use the methods outlined above to produce wavelet power spectra and identify significant features.  The results are then compared to the simulations in Section ~\ref {sim_section} and previous literature results.

\subsection{ARK 120}

An observation of the `bare' Seyfert 1 galaxy  Ark~120 ($z = 0.03271$) \citep{ark120paper} (Obs ID: 0147190101) is examined in the $0.3 - 10$ keV energy band.  The X-ray spectrum of Ark~120 is unremarkable, exhibiting a smooth soft excess above a harder power law, and a narrow Fe~K$\alpha$ emission line (\citealt{nardini2011reflection_ark120_softexcess}, \citealt{nandi2021long_ark120_softexcess}).  There is no observed warm absorber, hence it is labeled as `bare'. The central black hole has a high mass of $(150 \pm 19) \times 10^6 \, M_{\odot}$ \citep{peterson2004ark120_blackholemass}.

\begin{figure*}
    \centering
    \includegraphics[width=\columnwidth]{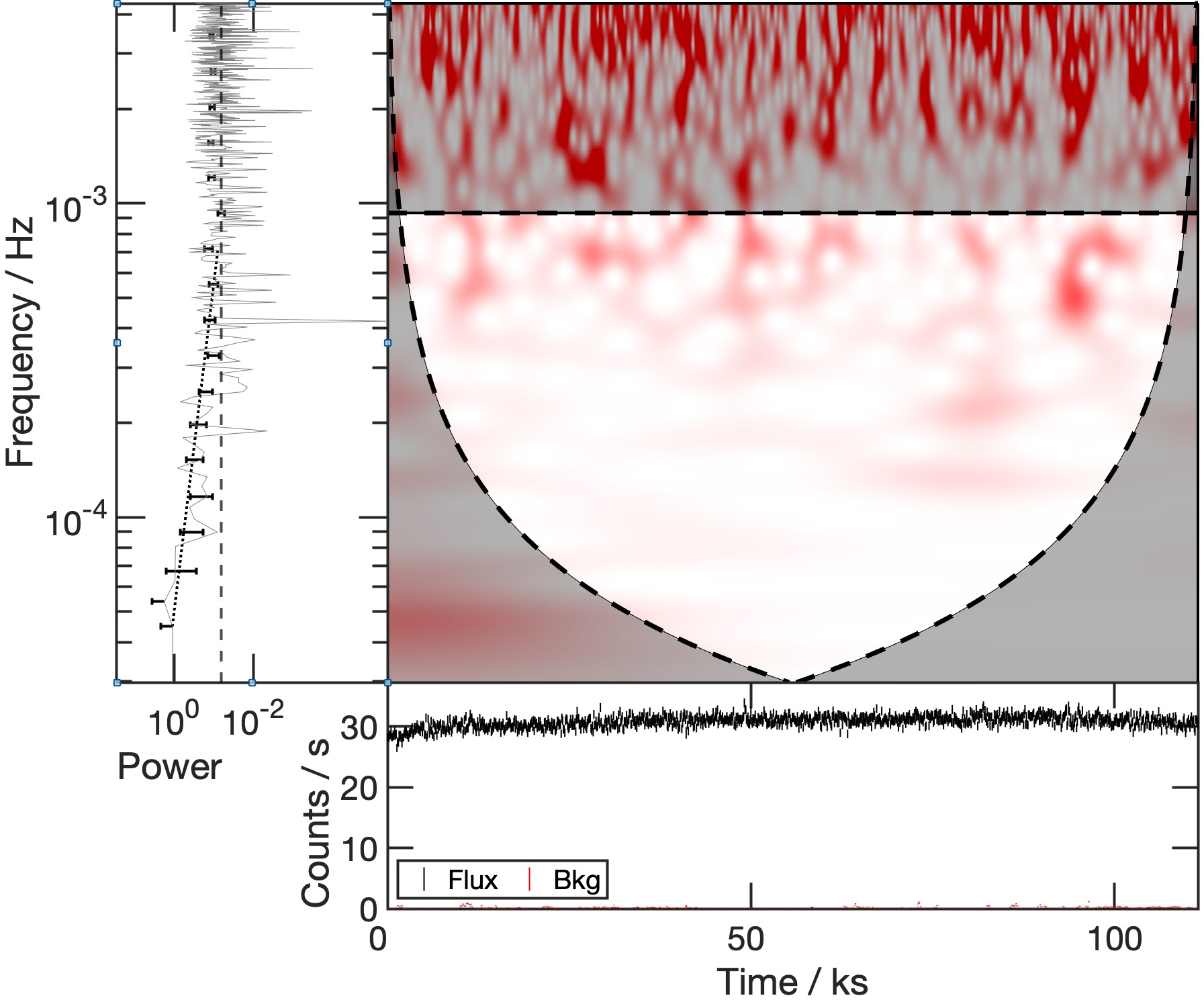}
    \includegraphics[width=\columnwidth]{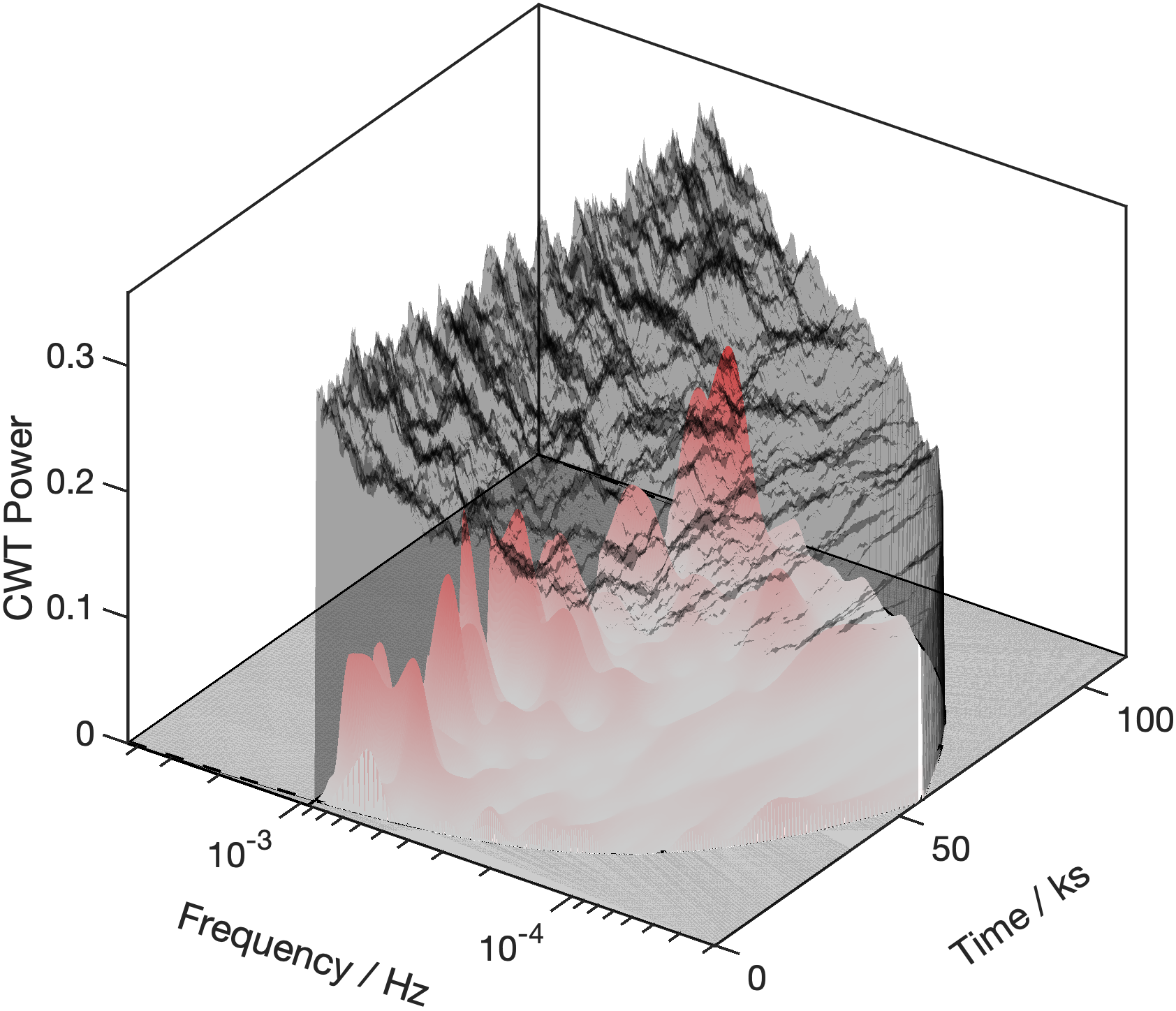}
    \caption{Left panel: The wavelet power spectrum of Ark~120 (Obs ID: 0147190101) appears similar to the white noise spectrum seen in Fig.~\ref{fig:noise_cwts}. Right panel: The wavelet power shown in 3-dimensions emphasizing there are no significant features.}
    \label{fig:ark_120_cwt}
\end{figure*}

Ark~120 serves as a good control test for our wavelet analysis.  The object is very bright ($\sim$ 30 counts per second), and shows low variability (lower panel of Fig.~\ref{fig:ark_120_cwt}), compared to most other Seyfert~1s.  This is consistent with work showing the excess variance, which is a measure of variability, is significantly (linearly) anti-correlated with black hole mass \citep{lu2001_excessvariance_BHmass}.  

The measured PSD is very flat ($\beta\approx 0.96$) at frequencies not dominated by the Poisson noise.  In Fig.~\ref{fig:ark_120_cwt}, the wavelet spectrum appears completely consistent with a flat noise spectrum (e.g. compare to top panel of Fig.~\ref{fig:noise_cwts}) with no obvious frequencies of importance. 


\subsection{IRAS 13224-3809}
IRAS~13224--3809 ($z = 0.066$) is a highly variable Narrow-Line Seyfert 1 (NLS1; e.g. \citealt{gallo2018_nls1_rapid}) active galaxy. This AGN is reported to have a maximum spinning black hole $a_* > 0.94$ \citep{jiang2018_1_iras_spin} and a black hole mass of $(1.9 \pm 0.2) \times 10^6 \, M_{\odot}$ measured from X-ray reverberation \citep{alston2020dynamic_iras}. The low black hole mass is expected to be accompanied with high flux variability.

If Ark~120 is noted for its simplicity, then IRAS~13224-3809 exemplifies the opposite extreme.  For decades, it has been recognized for its persistent, rapid, and high-amplitude variability (e.g. \citealt{boller1997_iras_rapid}; \citealt{dewangan2002_iras_variable}; \citealt{gallo2004_iras_variable}). Deep observations with \textit{XMM-Newton} show the source possesses a strong reverberation lag at a frequency of $\sim 2 - 5 \times 10^{-4}$ Hz (e.g. \citealt{fabian2013_iras_lag}, \citealt{kara2013_iras_reverberation_lag}).  There is also evidence of a QPO at $7 \times 10^{-4}$ Hz \citep{alston2019remarkable}.  In addition, there is evidence the AGN X-ray variability is non-stationary as IRAS~13224--3809 displays a time-dependent PSD, non-log-normal flux distribution, and a non-linear rms-flux relation \citep{alston2019remarkable}.

With the exceptional amount of \textit{XMM-Newton} data, IRAS~13224-3809 is an excellent target for a wavelet analysis.  The AGN provides the opportunity to investigate non-stationary processes and the persistence of QPOs and reverberation lags.  In this work, we examine only one observation (Obs ID: 	
0780561601) in the $0.3 - 5$ keV energy band, and leave analysis of all the data for future work.
\begin{figure*}
    \centering
    \includegraphics[width=\columnwidth]{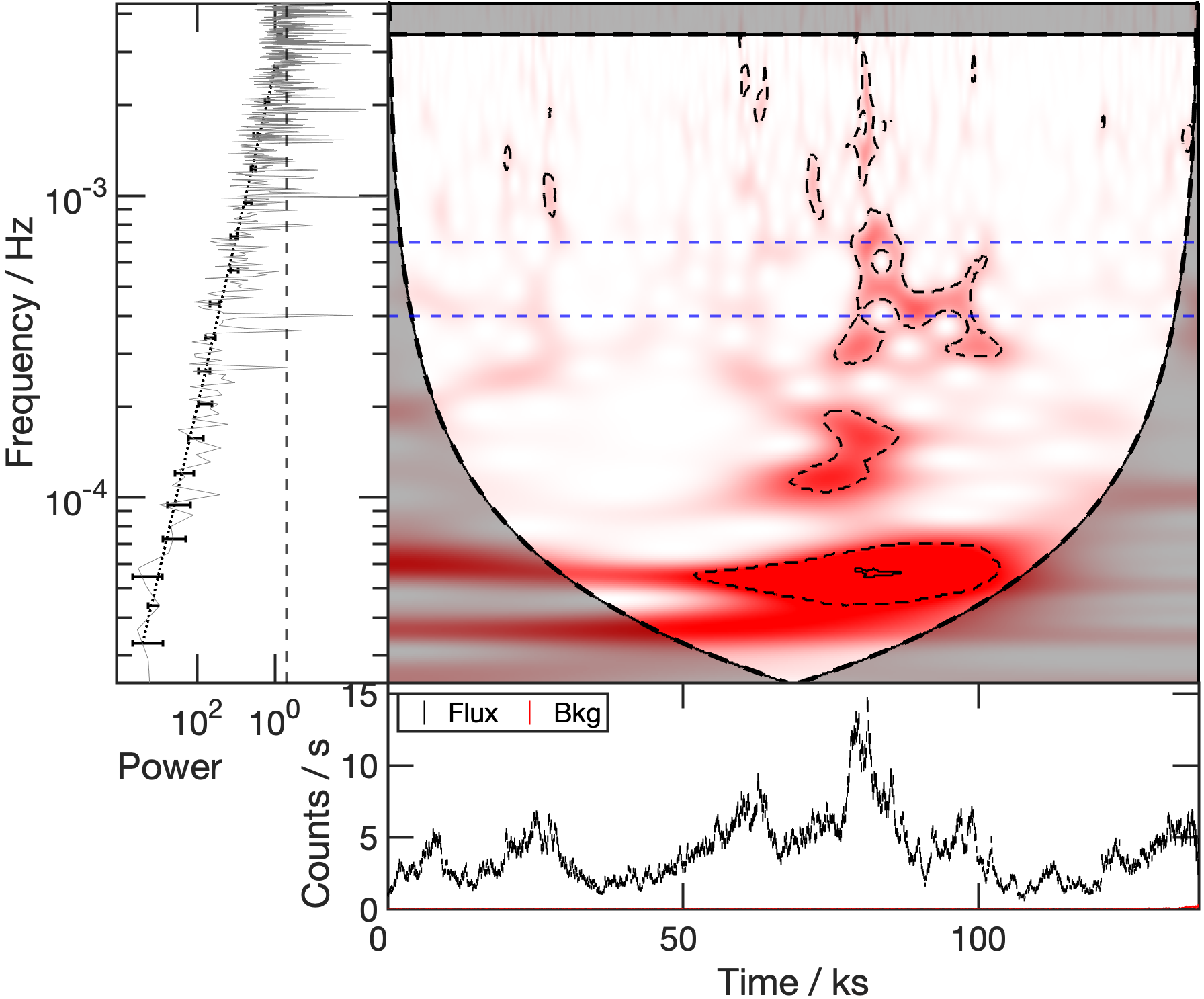}
    \includegraphics[width=\columnwidth]{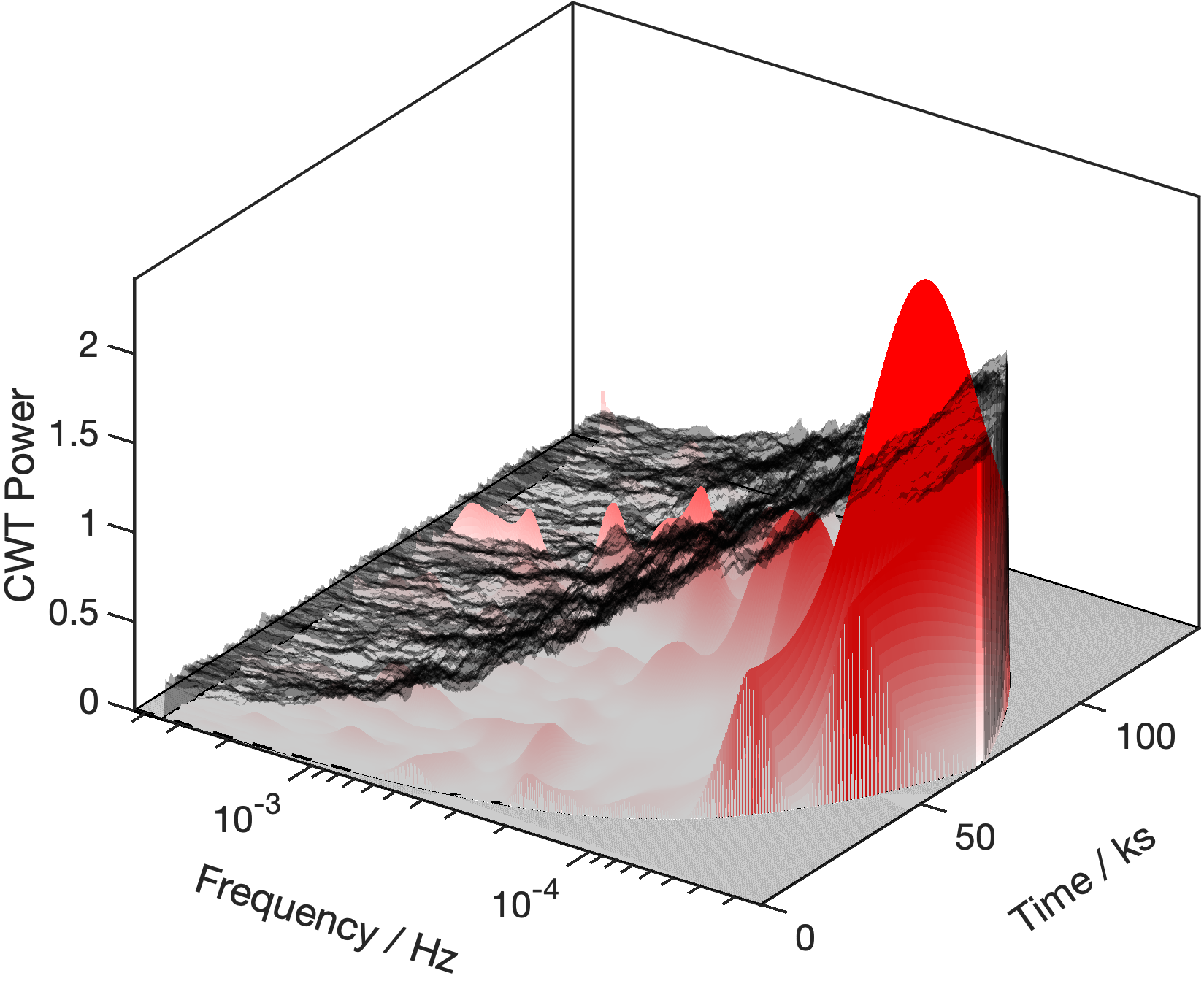}
    \caption{Wavelet power spectrum of IRAS 13224-3809 from revolution 3044 (Obs ID: 0780561601). Left panel: The higher frequency blue dashed line represents the QPO frequency at $7 \times 10^{-4}$ Hz \citep{alston2019remarkable}. The lower frequency blue dashed line represents the reverberation lag frequency at $\sim 4 \times 10^{-4}$ Hz \citep{kara2013_iras_reverberation_lag}.  Right panel: The amplitude peak coincides with the flare in the light curve. It is also apparent that for a steeper red noise PSD such as this one, that the plane of significance is also steep with respect to the frequency axis. Thus, amplitudes at lower frequencies must be much greater to be significant.}
    \label{fig:iras_cwt_double_panel}
\end{figure*}

The wavelet power spectrum for IRAS~13224-3809 in Fig. \ref{fig:iras_cwt_double_panel} is dominated by power at lower frequencies, less than $\sim10^{-4}$ Hz. In addition, during a flaring event at $\sim75$~ks, the wavelet power increases at all frequencies.  In Section~\ref{colours_of_noise_section}, we found that as $\beta$ increases, the wavelet amplitude increases toward lower frequencies.   This observation of  IRAS~13224-3809 was measured to have $\beta = 1.8$ and we can make comparisons with the steeper noise spectrum seen in Fig.~\ref{fig:noise_cwts}.  Moreover, in the noise simulation in Fig.~\ref{fig:noise_cwts}, we note that flares in the count rate can increase the wavelet power at all frequencies.   Looking specifically at the frequencies associated with the reverberation lag ($\sim 2 - 5 \times 10^{-4}$ Hz) and the QPO ($7 \times 10^{-4}$ Hz), we note that the power at these frequencies is only significant during the flaring event at $\sim75$~ks.

The wavelet spectrum of IRAS~13224-3809 is intriguing and motivates a deeper analysis of all the available data.  The single observation we have examined here displays behaviour that is consistent with simple red-noise fluctuations.  Analysis of all the available light curves would establish if there is some time or flux dependency for the QPO and reverberation signals.

\subsection{RE J1034+396}
\label{rej1034396}
QPOs are regularly observed in stellar-mass black holes (e.g. \citealt{remillard1999_stellarmass_qpo}, \citealt{strohmayer2001_stellarmass_qpo}, \citealt{remillard2002_stellarmass_qpo}, \citealt{remillard2006_xrb_stellarmass_qpo}, \citealt{vaughan2011rapid_stellarmass_qpo}), but the case is less robust in AGN.  The best example of a QPO in an AGN is from observations of the NLS1 galaxy,  RE J1034+396 $(z = 0.0424)$ \citep{gierlinski2008_rej_qpo}, which has an estimated mass of $\sim 2 \times 10^6 M_{\odot}$ \citep{middleton2009_rej_mass}. Repeated observations of RE  J1034+396 have confirmed the existence of the QPO, which appears to be caused by a dynamic structure in the inner disk \citep{jin2021rej}.  When present, the QPO occurs at $\sim 2.8 \times 10^{-4}$ Hz.

\begin{figure*}
    \centering
    \includegraphics[width=\columnwidth]{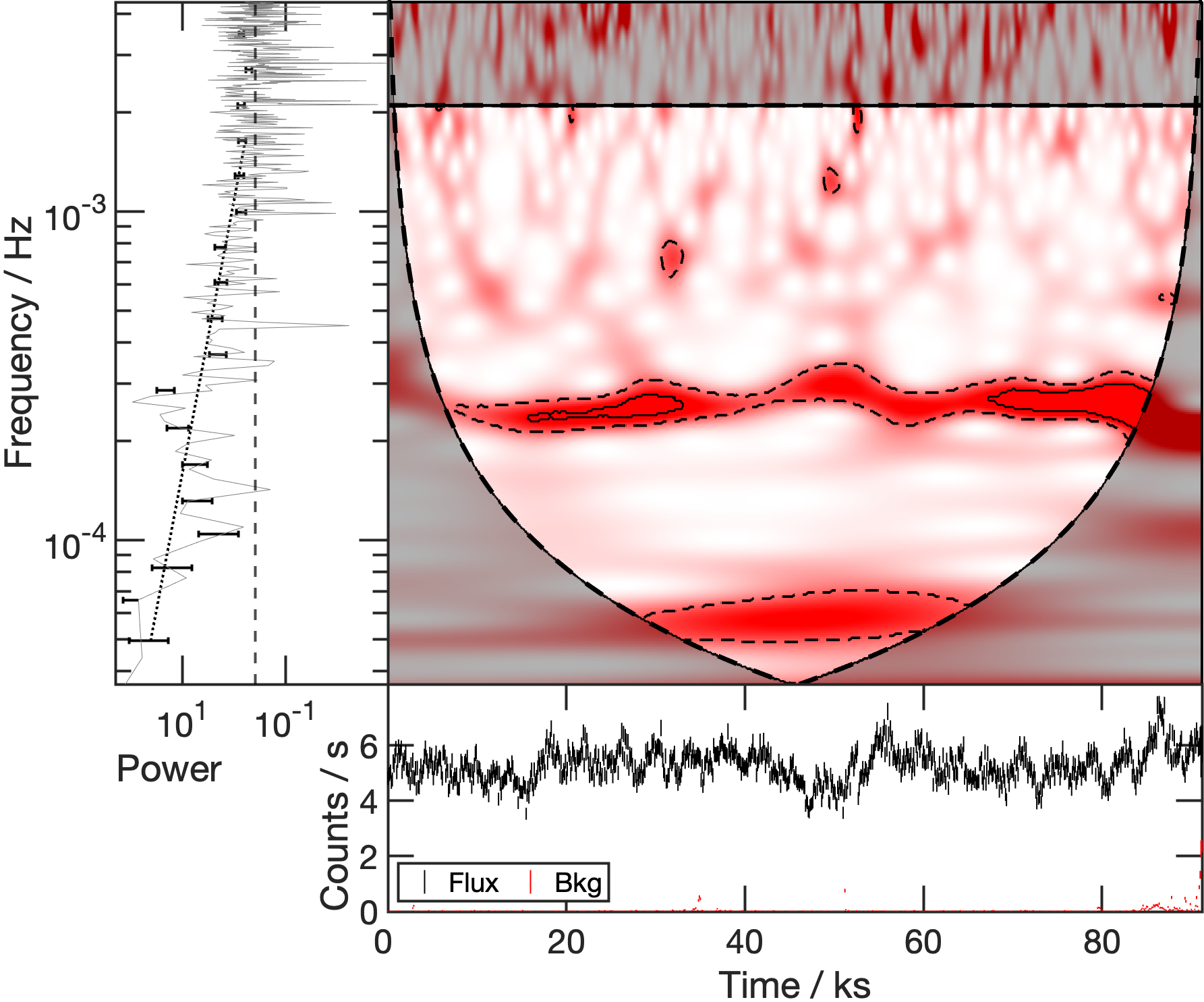}
    \includegraphics[width=\columnwidth]{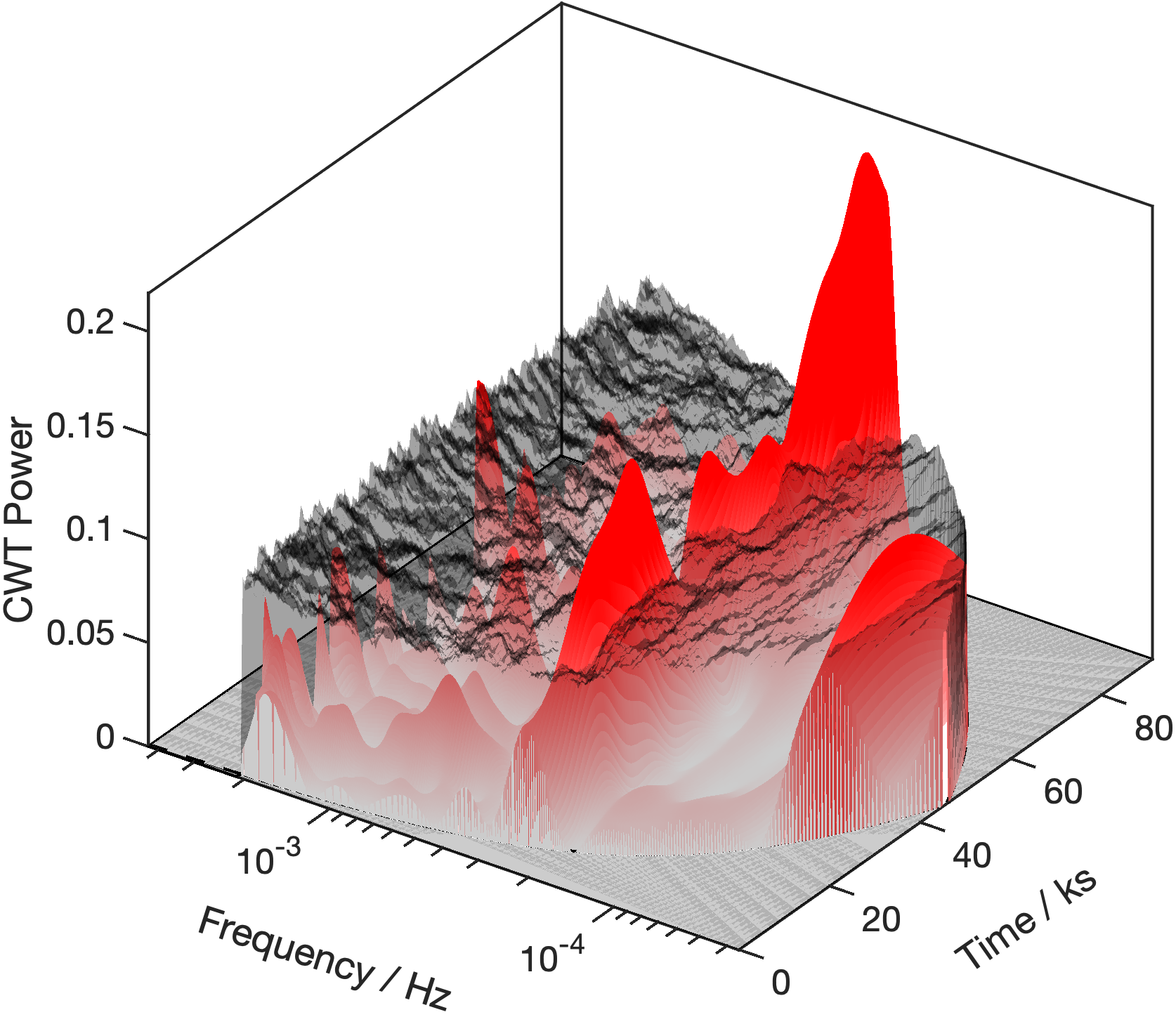}
    \caption{Wavelet power spectrum of RE J1034+396 (Obs ID: 0506440101). The strong, persistent QPO (e.g. \citealt{rej2020qpo}, \citealt{jin2021rej}) is apparent here. Left panel: It is apparent that the QPO is fully persistent and significant at the 90\% confidence level. It is not fully persistent at the 99\% confidence level. Right panel: We can see that the QPO is almost fully persistent at the 95\% confidence level. This 3-dimensional representation accentuates the variability of the Lorentzian component. There is also a significant frequency component at $\sim 6 \times 10^{-5}$ Hz, which is likely due to the underlying red noise.}
    \label{fig:rej 05 cwt double panel}
\end{figure*}

Examining the 2007 {\it XMM-Newton} observation (Obs ID: 0506440101) of RE J1034+396 in the $0.3 - 10$ keV energy band, which exhibited the strongest signal, finds the feature significantly detected in the wavelet power spectrum (Fig.~\ref{fig:rej 05 cwt double panel}).  At the 90\% confidence level it is persistent across the entire observation, but the signal is more sporadic at the 99\% confidence level. The signal compared to the 95\% confidence level is shown in the 3-dimensional representation (Fig. \ref{fig:rej 05 cwt double panel}). This 3-dimensional representation accentuates the variability of the QPO signal in the light curve.

The fluctuating significance of the feature does not necessarily indicate the signal is variable.  As seen in Section~\ref{sim_section}, the significance of a constant feature can fluctuate in a red noise spectrum or if the underlying noise is non-stationary (Fig.~ \ref{fig:simulated_nonst_pl_const_qpo}).  In addition, if the physical signal resembles a Lorentzian profile with some width in frequency, it can also exhibit fluctuations in wavelet power (Fig.~\ref{fig:sin_waves_cwt_for_coi_eg}). 

There is also a feature at the 90\% confidence level at $\sim 
6 \times 10^{-5}$ Hz.  As in Fig.~\ref{fig:noise_cwts}, such a feature could be attributed to the steep underlying power spectrum ($\beta = 1.2$) in RE  J1034+396.

\subsection{GSN 069}


GSN 069 ($z = 0.018$) was the first galaxy discovered to show QPEs in its X-ray flux \citep{miniutti2019nine_gsn}. These eruptions are short-lived X-ray flares with a duration of approximately 1 hour occurring approximately every 9 hours. The QPEs are transient in GSN 069 (e.g. \citealt{miniutti2023_gsn}) and several scenarios have been proposed to describe the behaviour, including unstable mass transfer occurring between a low mass star and the supermassive black hole \citep{lu2022_gsn_qpe_unstable_mass_transfer}, or instabilities in the inner accretion disk \citep{pan2022_gsn069_disk}.   The behaviour could be linked to a previous tidal disruption event (TDE) in the galaxy (e.g. \citealt{miniutti2019nine_gsn}, \citealt{miniutti2023_gsn}).  The black hole mass has been estimated to be on the order of $10^5 \, M_{\odot}$ \citep{miniutti2023_gsn}. 

\begin{figure*}
    \centering
    \includegraphics[width=\columnwidth]{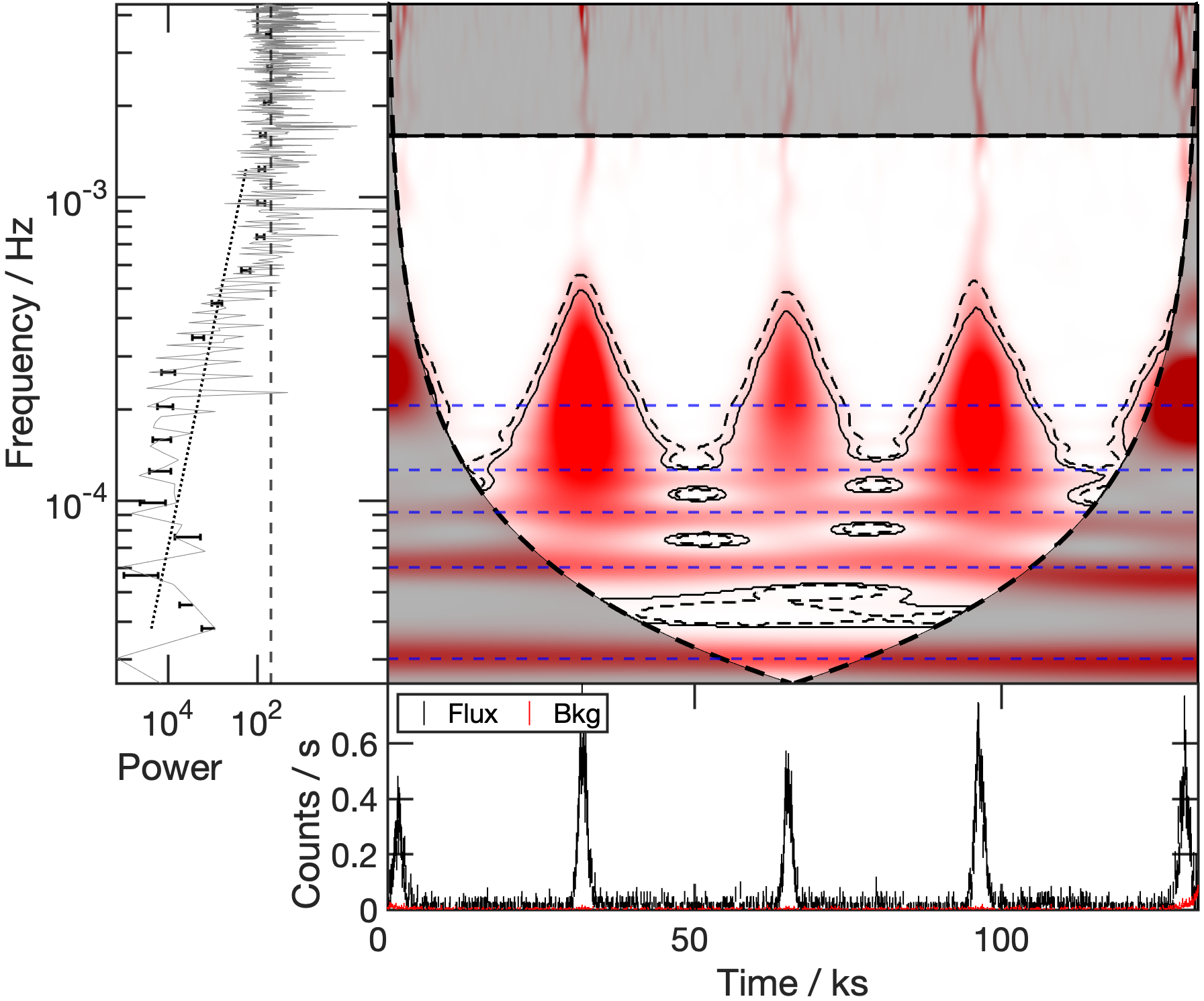}
    \includegraphics[width=\columnwidth]{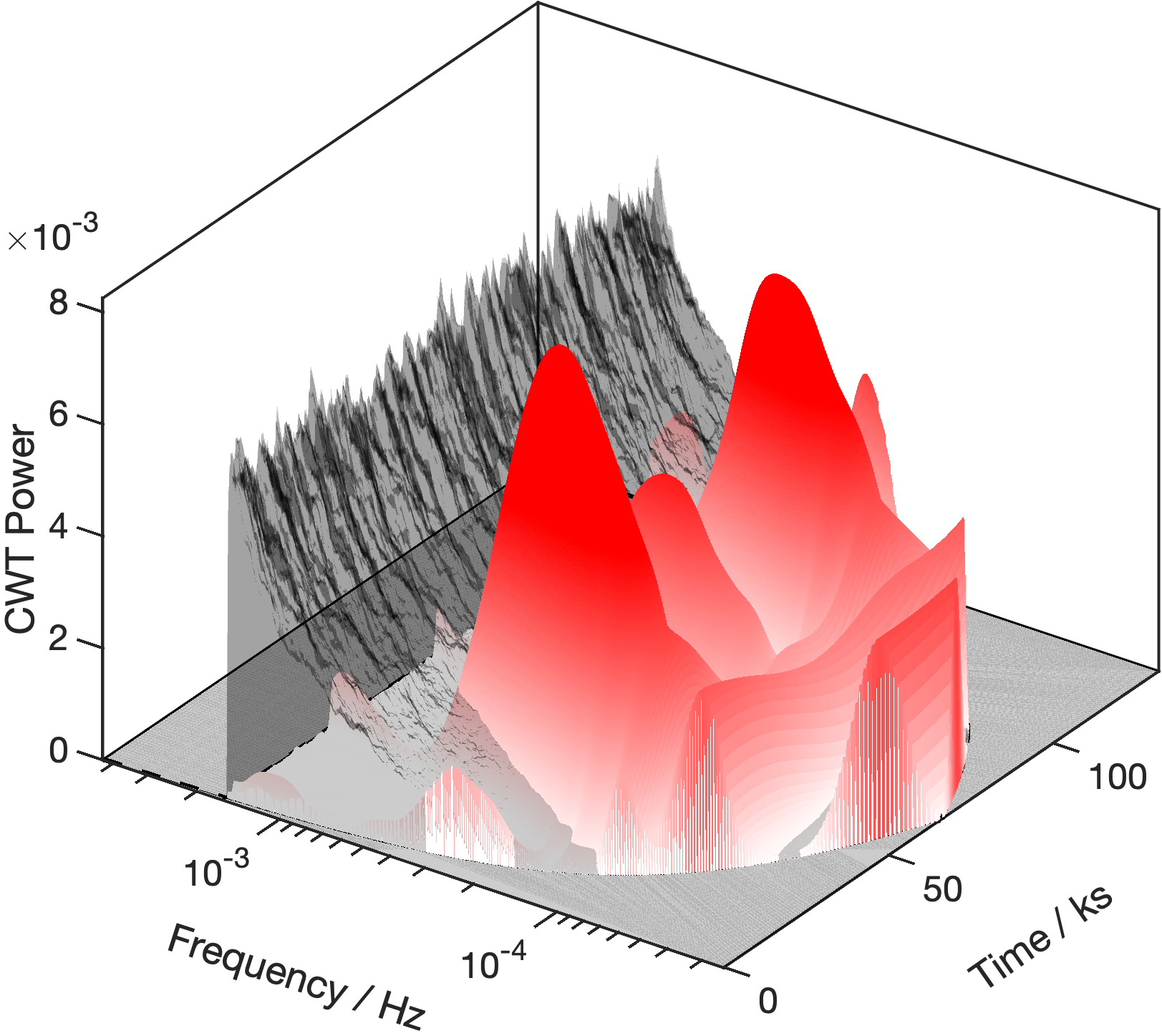}
    \caption{The wavelet power spectrum of GSN 069 (Obs ID: 0851180401). Left panel: The dashed blue lies represent the harmonics, which are taken to be the frequencies corresponding to peaks in the global wavelet power spectrum (Fig. \ref{fig:gsn_gwt}. Right panel: The 3-dimensional representation emphasizes the amplitude peaks in the vicinity of the eruption events in this light curve.}
    \label{fig:gsn_cwt}
\end{figure*}

The observation of GSN 069 used here is from May 2019 (Obs ID: 0851180401) in the $0.5 - 2$ keV energy band, and has an exposure of 132 ks.  The wavelet power spectrum of  GSN 069 might be the most unusual one we have examined here (Fig. \ref{fig:gsn_cwt}).  The wavelet power exhibits amplitude peaks across a range of frequencies during each of the eruption events in the light curve. However, there are also a number of persistent frequencies throughout the light curve.   

\begin{figure}
    \centering
    \includegraphics[width=\columnwidth]{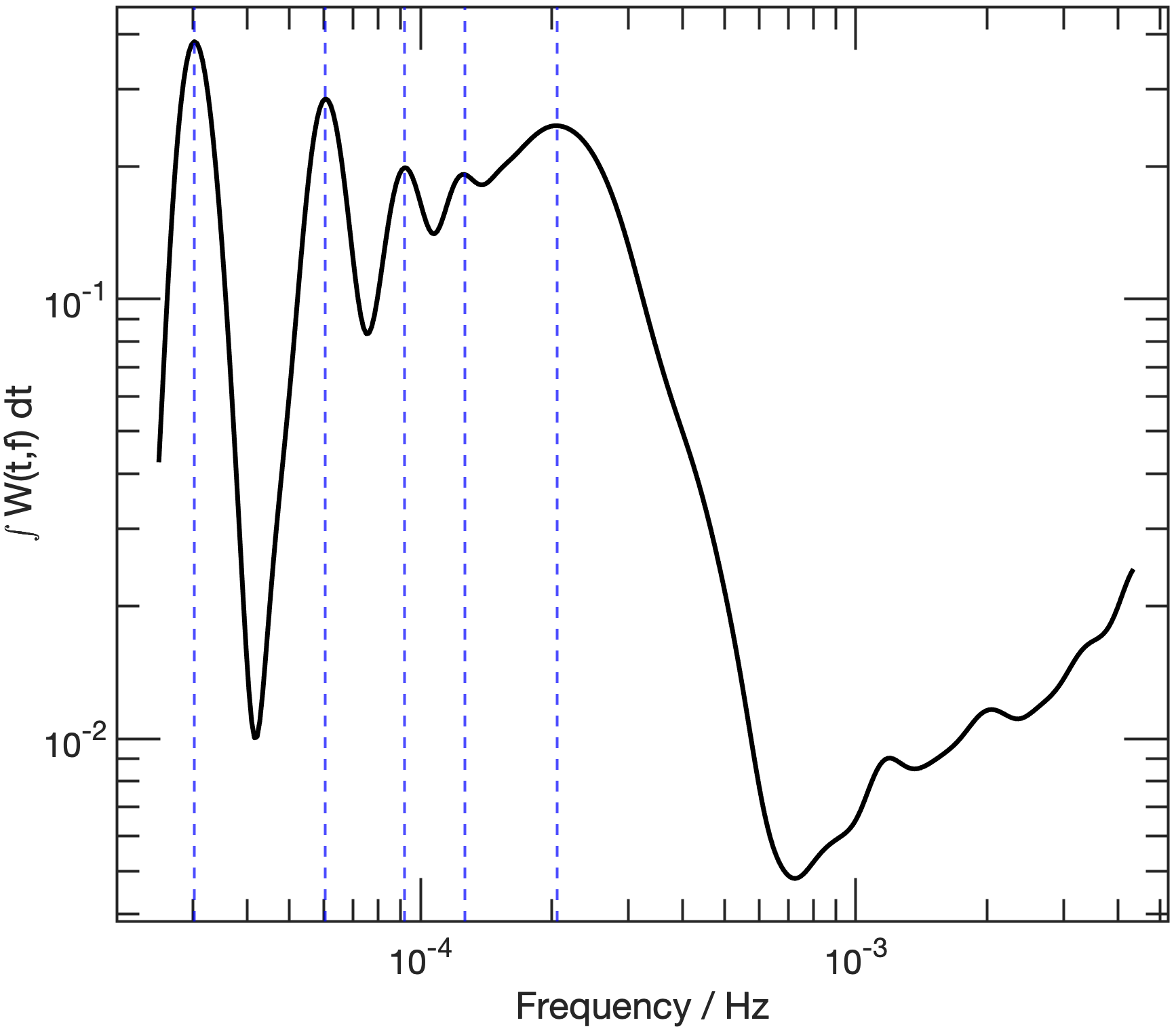}
    \caption{Global wavelet transform of GSN 069.  The vertical lines highlight the frequencies where the wavelet power peaks.  These will naturally coincide with the most significant regions in Fig.~\ref{fig:gsn_cwt}.}
    \label{fig:gsn_gwt}
\end{figure}

Comparing the wavelet power spectrum to the simulations in Section~\ref{sim_section}, we note a striking resemblance with Fig.~\ref{fig:simulated_qpe} that displays harmonic behaviour over a white noise spectrum.  We can further examine the wavelet power by calculating the global wavelet transform (Fig.~\ref{fig:gsn_gwt}), which effectively integrates the total power at each frequency over all times (as defined in \citealt{torrence1998}).  There are five peaks that obviously coincide with the statistically significant contours in Fig. \ref{fig:gsn_cwt}.  These peaks are at frequencies of approximately 3.02, 6.04, 9.19, 12.62, and $20.62\times10^{-5}$~Hz.  

With the exception of the lowest frequency at $3.02\times10^{-5}$~Hz, which is consistent with the 9 hour flaring period and is sampled for the least amount of time in the light curve (just at the bottom of the COI), the ratio between each subsequent higher frequency is approximately 1.5 (specifically $\sim 1.5, 1.4, \text{and } 1.6$).  The 3:2 ratio is coincident with what is observed in the QPOs of X-ray binaries.

However, we cannot rule out that this pattern is an artefact of the the underlying Fast Fourier Transform process present within the calculation of the CWT. Simulations have been done on an artificial signal consisting of a series of evenly spaced delta functions. The CWT was calculated on these signals, which also exhibited a similar pattern between the peaks of the GWT. When dividing the frequency of the $n^{th} + 1$ harmonic by the $n^{th}$ harmonic, the ratios followed as 2/1, 3/2 and 4/3, after which the GWT was dominated by noise. The two latter ratios being slightly different for observations such GSN 069 could be due to the QPE peaks not being exactly evenly spaced, like with the delta function signal. For reference, the three lowest frequency ratios of the harmonics present in GSN 069 were 2.0, (3/2 + 0.02) and (4/3 + 0.04).

\section{Discussion and Conclusions}


With the understanding that the X-ray emission from AGN may be non-stationary, it is important to consider timing analysis tools that account for this. Fourier transforms which are widely used for analyzing AGN variability make the underlying assumption that the time series is stationary. The wavelet transform makes no such assumption.  In addition, the wavelet transform preserves timing information while extracting the important frequencies in a series. In this work, we examine the application of the wavelet transform to the X-ray timing properties of AGN and QPEs.


Several simulations are carried out to test the effectiveness of the wavelet analysis for AGN light curves.   The most basic was to examine the effects of different colours of underlying noise (i.e. different values of $\beta$ for the PSD). As the slope increases, more noise will be introduced in the wavelet image at lower frequencies (Fig.~\ref{fig:noise_cwts}).  This needs to be considered for objects with steep PSDs like IRAS~13224-3809, which displays a persistent-looking feature at low frequencies (e.g. Fig. \ref{fig:iras_cwt_double_panel}) that might simply be noise.


Since overall flux can vary significantly in AGN, we tested the effects of different average count rates and exposure lengths for wavelet analysis (e.g. Fig.~\ref{fig:cwt_changing_avg_count_rate}).  The main effect of the observation duration is the ability to explore lower frequencies and detect signals with higher significance.  Obviously, the more often a frequency is observed in a light curve the more confidence there is in its importance.  

The brightness of the AGN is less important at detecting signals in wavelets than the relative strength of the signal above the underlying PSD.  A strong signal can be significantly detected at low count rates.  It is obviously important that there is sufficient data to get a reasonable representation of the PSD and this is easier to achieve with bright objects (e.g. Fig.~\ref{fig:different_L_PL_ratios}).  Signals that exhibit a Lorentzian-to-power law ratio of $\sim 10\%$ can be robustly detected.  The signal strength is comparable to that of the QPO in RE J1034+396.


The manner in which the periodic signal is modelled can also influence the appearance of the wavelet power spectrum.  QPOs in AGN and X-ray binaries are often modelled with Lorentzian profiles.  In this work we considered pure sine waves as well as Lorentzian profiles (Fig.~\ref{fig:sin_waves_cwt_for_coi_eg}).  A sine function results in a persistent feature in the wavelet spectrum that is at constant power.  The situation is more complicated with the Lorentzian profile as it possesses some natural width (i.e. range of frequencies).  The power in the feature can appear to fluctuate with time in the wavelet spectrum even if it is persistent.

This is important to keep in mind when examining the time dependency of a periodic signal.  For example, the QPO in RE J1034+396 appears persistent over the duration of the observation, but its power appears to fluctuate.  This might simply arise from the nature of the Lorentzian signal (Fig.~\ref{fig:rej 05 cwt double panel}).


In considering the variable nature of the Lorentzian profile we have touched on non-stationarity.  As part of this examination, we looked at the effects of a variable PSD on the ability to detect signals (Fig.~\ref{fig:simulated_nonst_pl_const_qpo}).   We find that periodic signals can still be recovered when modelling the variable underlying PSD with an average value. 


Wavelet power spectra were used to analyze an initial sample of four AGN, all exhibiting different behaviours from one another. For this work, the generalized Morse wavelet basis vector was used. By varying the time-bandwidth parameter, either the time-localized frequencies or persistent frequencies can be accentuated. For this work, the time-bandwidth parameter was held constant at the 60, which is provides moderate time and frequency resolution without preferring one over the other.

The wavelet power spectrum of Ark 120 was entirely consistent with noise. Given the large black hole mass and the low amplitude of variability, the result was expected. 

IRAS 13224-3809 exhibits a steep red noise PSD.  As stated, this will produce noise at lower frequencies in the wavelet.  For the most part, there are no discerning features in the wavelet, except during a count rate flare at $\sim 75$~ks (Fig.~\ref{fig:iras_cwt_double_panel}).  At that time, a number of frequencies are detected at $>90$ per cent confidence.  Interestingly, two of these frequencies correspond to the reverberation lag between $\sim 2 - 5 \times 10^{-4}$ Hz (e.g. \citealt{fabian2013_iras_lag, kara2013_iras_reverberation_lag, alston2019remarkable}) and the QPO at $7 \times 10^{-4}$ Hz \citep{alston2019remarkable} previously reported in this object.  This raises the possibility that these frequencies are time- and/or flux-dependent.  However, with only the one observation examined here, we cannot rule out that the enhanced power at all frequencies during the flare is not simply a red noise characteristic.  Analysis of the all the archival data of IRAS~13224-3809 will be important.

Examining the \textit{XMM-Newton} observation of RE J1034+396 that exhibited the strongest QPO, we find the feature is significantly detected in the wavelet power spectrum (Fig.~\ref{fig:rej 05 cwt double panel}).  At the 90\% confidence level it is persistent across the entire observation, but the signal is more sporadic at the 99\% confidence level. The fluctuations in the power of the QPO could be attributed to the Lorentzian nature of the signal.

GSN 069 possesses a unique wavelet power spectrum compared to the other sources examined, as it is a rare AGN that exhibits QPE phenomena in its light curves. There are several frequencies that are significantly detected.  The primary frequency at $\sim 3\times10^{-5}$~Hz is coincident with the $\sim9$~hour flares.  However, there are higher frequencies detected with significance at 6.04, 9.19, 12.62, and $20.62\times10^{-5}$~Hz.  The ratio between these frequencies and each subsequent lower frequency is approximately 1.5 exhibiting a roughly 3:2 ratio that is similar with what is observed in the QPOs of X-ray binaries.
This pattern cannot yet be rules out as being an artefact of the CWT calculation. The origins of QPOs is still uncertain, but this coincidence in the ratio of the oscillation frequencies draws direct comparison between supermassive and stellar mass black holes.

Wavelet analysis is a potentially powerful tool to study AGN X-ray light curves. It is complimentary to Fourier analysis and provides the potential to reveal signals that are transient in time or variable in power.  Deeper analyses making use of multi-epoch data to examine transient features in the wavelet power spectrum of AGN, X-ray binaries and QPE sources will be important.

\section*{Acknowledgements}

This work was based on observations obtained with XMM-Newton, an ESA science mission with instruments and contributions directly funded by ESA Member States and NASA. This research has made use of the NASA/IPAC Extragalactic Database (NED), which is funded by the National Aeronautics and Space Administration and operated by the California Institute of Technology. LCG acknowledge financial support from the Natural Sciences and Engineering Research Council of Canada (NSERC) and from the Canadian Space Agency (CSA).

\section*{Data Availability}

The data used for this study is available on the \textit{XMM-Newton} Science Archive (XSA).




\bibliographystyle{mnras}
\bibliography{sources}








\newpage

\bsp	
\label{lastpage}
\end{document}